\documentclass[11pt,english]{article}
\usepackage[latin9]{inputenc}
\usepackage{amsmath}
\usepackage{amssymb}
\usepackage{graphicx}
\usepackage{setspace}
\makeatletter
\@ifundefined{date}{}{\date{}}
\usepackage{esint}
\usepackage{hyperref}
\usepackage{latexsym}
\usepackage{graphicx}\usepackage{bm}\usepackage{longtable}

\usepackage{xcolor}

\@addtoreset{equation}{section}

\makeatother

\usepackage{babel}
\begin{document}
\title{Out-of-time-order correlator as a detector of baryonic phase structure
in holographic QCD with instanton}
\maketitle
\begin{center}
Si-wen Li\footnote{Email: siwenli@dlmu.edu.cn}, Yi-peng Zhang\footnote{Email: ypmahler111@dlmu.edu.cn},
Hao-qian Li\footnote{Email: lihaoqian@dlmu.edu.cn},
\par\end{center}

\begin{center}
\emph{Department of Physics, School of Science,}\\
\emph{Dalian Maritime University, }\\
\emph{Dalian 116026, China}\\
\par\end{center}

\vspace{12mm}

\begin{abstract}
We study the out-of-time-order correlators (OTOC) of Skyrmion as baryon
in the D0-D4/D8 model which is expected to be holographically dual
to QCD with instantons as D0-branes or with a non-zero theta angle.
Baryon states are identified to the excitations of the Skyrmion which
are described by a holographic quantum mechanical system in this model.
By employing the definition of OTOC in quantum mechanics, we derive
the formulas and demonstrate explicitly the numerical calculations
of the OTOC. Our calculation illustrates the quantum OTOC with imaginary
Lyapunov coefficient indicates the possibly metastable baryonic status
in the presence of the instanton while the classical OTOC can not,
thus it reveals the instantonic or theta-dependent features of QCD
are dominated basically by its quantum properties. Furthermore, the
OTOC also implies the baryonic phase becomes really chaotic with real
Lyapunov exponent if the instanton charge increases sufficiently which
agrees with the unstable baryon spectrum presented in this model.
In this sense, we believe the OTOC may be treated as a tool to detect
the baryonic phase structure of QCD.
\end{abstract}
\newpage{}

\section{Introduction}

In recent years, the out-of-time-order correlator (OTOC) has been
considered as a measure of the magnitude of quantum chaos \cite{key-1}
while it was first introduced to calculate the vertex correction of
a current for a superconductor \cite{key-2}. In general, the quantum
OTOC $C_{T}\left(t\right)$ is defined by using the commutator of
two operators as,

\begin{equation}
C_{T}\left(t\right)=-\left\langle \left[W\left(t\right),V\left(0\right)\right]^{2}\right\rangle ,
\end{equation}
where $W\left(t\right),V\left(t\right)$ are operators in the Heisenberg
picture at time $t$ and $\left\langle ...\right\rangle $ refers
to the thermal average. In particular, the authors of \cite{key-3,key-4,key-5}
suggest that the operators $W\left(t\right),V\left(t\right)$ in quantum
mechanics can be chosen as the canonical coordinate $x\left(t\right)$
and momentum $p\left(t\right)$. In this sense, the classical limit
of the OTOC can also be evaluated quantitatively if the quantum commutator
$\frac{1}{i}\left[,\right]$ is replaced by the Poisson bracket $\left\{ ,\right\} _{\mathrm{P.B}}$.
Accordingly, the chaos of classically mechanical system can be evaluated
through the OTOC as,

\begin{equation}
C_{T}\sim\left\{ x\left(t\right),p\left(0\right)\right\} _{\mathrm{P.B}}^{2}\sim\left[\frac{\delta x\left(t\right)}{\delta x\left(0\right)}\right]^{2}\sim e^{2Lt},\label{eq:2}
\end{equation}
where $L$ is the Lyapunov coefficient. For a chaotic system, $L$
must be positive and real since such a system is very sensitive to
the initial condition $x\left(0\right)$ otherwise it is not really
chaotic. However, the quantum version of the OTOC (\ref{eq:2}) could
instead saturate at the Ehrenfest time $t_{E}$ which refers to a
time scale that the wave function spreads over the whole system characterizing
the boundary between the particle-like and the wave-like behavior
of a wave function. Therefore, the exponent growth of the OTOC is
conjectured to be the characteristic property of the classical or
non-integrable system.

Moreover, the OTOC is also an important observable in the context
of quantum gravity, AdS/CFT correspondence or gauge-gravity duality
\cite{key-6,key-7}. In the context of quantum information on the
horizon of black hole \cite{key-8,key-9,key-10,key-11,key-12,key-13},
it implies an upper bound of the quantum Lyapunov coefficient as $L\leq2\pi T$.
On the other hand, the analysis of chaos in the Sachdev-Ye-Kitaev
(SYK) model \cite{key-14,key-15} (a quantum mechanical system with
infinitely long range disorder interactions of Majorana fermions)
illustrates the bound of quantum Lyapunov exponent is saturated \cite{key-16,key-17,key-18}.
Therefore, it strongly implies a quantum black hole can be described
by the SYK model in holography if the OTOC could be a tool to detect
the AdS/CFT correspondence or gauge-gravity duality.

Motivated by these, in this work we would like to introduce OTOC as
a tool to detect the baryonic phase structure of QCD in holography,
in order to further explore the role of OTOC in gauge-gravity duality.
To this goal, we aim at the D0-D4/D8 model \cite{key-19,key-20,key-21,key-22}
as an extension of the famous D4/D8 model (Witten-Sakai-Sugimoto model)
\cite{key-23,key-24,key-25,key-26,key-27} based on the IIA string
theory which is holographically dual to theta-dependent QCD. The reasons
to choose the D0-D4/D8 model are given as follows\footnote{While the experimental value of the theta angle in QCD may be small,
it attracts great interests to study some relevant effects with the
theta angle, e.g. the deconfinement phase transition \cite{key-30,key-31},
the glueball spectrum \cite{key-32}, the large N behavior of gauge
theory \cite{key-33}, chiral magnetic effect \cite{key-34,key-35}.
The details of the theta-dependent QCD in theory can be reviewed in
\cite{key-36}.}: First, the baryon sector in this model is a quantum mechanical system
which is totally analytical and simple enough in the strong coupling
region, hence the definition of OTOC in quantum mechanics is valid.
Second, as a holographic description, baryon sector in the presence
of a non-vanished theta angle or instantons as D0-branes would include
various phase structures in this model \cite{key-28,key-29}, since
in the framework of QFT (quantum field theory), QCD vacuum includes
various phase structures with instanton as well \cite{key-R1,key-R2,key-R3}.
Therefore we expect the OTOC may somehow detect the these phase structures
through Lyapunov exponent in holography. Keep these in hand, our calculation
illustrates that the quantum OTOC does not grow exponentially when
the theta angle or instanton charge is sufficiently small, instead
it trends to become oscillated periodically in the large $N_{c}$
limit, and the period increases by the theta angle or instanton charge.
So if we write the OTOC as an imaginary exponent as it is suggested
in \cite{key-4}, the Lyapunov coefficient indicates the possible
baryonic status with non-zero theta angle or instanton charge in QCD
which is recognized to be metastable in this model \cite{key-21,key-22,key-28,key-29}.
However, the analysis of the classical OTOC does not lead to this
conclusion sensibly. Besides, our analysis also implies the real Lyapunov
exponent may arise in the OTOC at large time when the instanton charge
increases greatly, therefore it implies the baryonic phase becomes
unstable and really chaotic with sufficiently large instanton charge
or theta angle. In this sense, the holographic OTOC as a tool seems
possible to detect the baryonic phase structure of QCD and implies
that the features of QCD in the presence of the instanton or theta
angle is dominated basically by the quantum properties of the theory
as it is discussed in the framework of QFT \cite{key-R1,key-R2,key-R3}.

The outline of this paper is as follows. In Section 2, we collect
the essential parts of the Skyrmion as baryon in the D0-D4/D8 model
as a quantum mechanical system, then derive briefly the formulas of
the OTOC. In Section 3, we illustrate our numerical evaluation of
the OTOCs with various parameters and analyze their behaviors. In
Section 4, the classical limit of the OTOCs is discussed. The final
section as Section 5 gives the summary. In addition, the appendix
includes some essential calculations in this work.

\section{The holographic setup }

\subsection{The holographic Skyrmion in the D0-D4/D8 model}

The D0-D4/D8 model\footnote{The details of this model can be reviewed in \cite{key-19,key-20,key-21,key-22}.}
is a top-down holographic approach to QCD with instantons or a non-vanished
theta angle which is also an extension of the D4/D8 model (the Witten-Sakai-Sugimoto
model) \cite{key-23,key-24,key-25,key-26,key-27}. The Skyrmion as
baryon in this model is recognized as the collective mode of a baryon
vertex which is a D4-brane wrapped on the spherical part $S^{4}$
of the bulk geometry \cite{key-37,key-38} and it can be equivalently
described by the instanton configuration of the gauge field on the
flavor brane \cite{key-39,key-40,key-41}. By evaluating the mass
of the instantonic soliton on the flavor brane, the quantized Hamiltonian
in moduli space for Skyrmion as baryon in the D0-D4/D8 model is collected
as \cite{key-28},

\begin{align}
H= & M_{0}+H_{Z}+H_{y},\nonumber \\
H_{Z}= & -\frac{1}{2m_{Z}}\partial_{Z}^{2}+\frac{1}{2}m_{Z}\omega_{Z}^{2}Z^{2},\nonumber \\
H_{y}= & -\frac{1}{2m_{\rho}}\sum_{I=1}^{n+1}\frac{\partial^{2}}{\partial y_{I}^{2}}+\frac{1}{2}m_{\rho}\omega_{\rho}^{2}\rho^{2}+\frac{K}{m_{\rho}\rho^{2}},\label{eq:3}
\end{align}
where $a=\frac{1}{216\pi^{3}}$ and

\begin{align}
M_{0}= & 8\pi^{2}\lambda ab^{3/2}N_{c},m_{Z}=\frac{1}{2}m_{\rho}=8\pi^{2}ab^{3/2}N_{c},\nonumber \\
\omega_{Z}^{2}= & \frac{1}{3}\left(3-b\right),\omega_{\rho}^{2}=\frac{1}{12}\left(3-b\right),K=\frac{2}{5}N_{c}^{2}.\label{eq:4}
\end{align}
Here $N_{c},N_{f}$ refers respectively to the color and flavor number
(i.e. numbers of D4-branes and D8-branes). $\lambda$ is the 't Hooft
coupling constant. $b$ relates to the charge density of the D0-branes
as the bulk instanton, and for a given branch, $b$ also relates to
the theta angle of QCD in holography by \cite{key-20,key-21,key-22,key-42},

\begin{equation}
\theta\propto\sqrt{\frac{b-1}{b}}.\label{eq:5}
\end{equation}
Thus we have $1\leq b<3$ which implies the baryonic phase is possible
to be stable when the theta angle is sufficiently small since $b>3$
means $\omega_{Z},\omega_{\rho}$ are all imaginary so that the Skyrmion
as baryon is totally unstable. $n$ relates to the flavor number $N_{f}$
of the Skyrmion through the numbers of the $SU\left(N_{f}\right)$
generators as $n=N_{f}^{2}-1$. We note that the above formulas are
written in the unit of $M_{KK}=1$. $\rho$ refers to the size of
the instanton which denotes the radial coordinate in the moduli space
parameterized by $y_{I}$ as $\rho^{2}=\sum_{I}y_{I}^{2}$.

The eigen equation with respect to Hamiltonian (\ref{eq:3}) as,

\begin{equation}
H\psi=E\psi,
\end{equation}
can be solved analytically. The associated eigenfunction $\psi$ is
given as,

\begin{equation}
\psi_{n_{Z},n_{\rho},l_{n},l_{n-1},...l_{1}}\left(Z,\rho,\theta_{n},\theta_{n-1},...\theta_{1}\right)=\chi_{n_{Z}}\left(Z\right)\mathcal{R}_{n_{\rho},l_{n}}\left(\rho\right)\mathcal{Y}\left(\theta_{n},\theta_{n-1},...\theta_{1}\right),\label{eq:7}
\end{equation}
where 
\begin{align}
\mathcal{R}_{n_{\rho},l_{n}}\left(\rho\right) & =\mathcal{N}\left(n_{\rho},l_{n}\right)e^{-\frac{m_{\rho}\omega_{\rho}\rho^{2}}{2}}\rho^{\xi}F\left(-n_{\rho},\xi+\frac{n+1}{2},m_{\rho}\omega_{\rho}\rho^{2}\right),\nonumber \\
\xi & =\sqrt{2K+\left(l_{n}+\frac{n-1}{2}\right)^{2}}-\frac{n-1}{2},
\end{align}
is given by the confluent hypergeometric function $F\left(a,b,x\right)$
satisfying the normalization

\begin{equation}
\int d\rho\rho^{n}\mathcal{R}_{n_{\rho}^{\prime},l_{n}^{\prime}}\mathcal{R}_{n_{\rho},l_{n}}=\delta_{n_{\rho}^{\prime},n_{\rho}}.
\end{equation}
And
\begin{equation}
\mathcal{Y}\left(\theta_{n},\theta_{n-1},...\theta_{1}\right)=\frac{\left(-1\right)^{l_{1}}}{\sqrt{2\pi}}e^{il_{1}\theta_{1}}\prod_{j=2}^{n}\ _{j}\mathcal{P}_{l_{j}}^{l_{j-1}}\left(\theta_{j}\right),\label{eq:10}
\end{equation}
 is the spherical harmonic function on $S^{n}$. We note that $\ _{j}\mathcal{P}_{l_{j}}^{l_{j-1}}\left(\theta_{j}\right)$
is given by combining the associated Legendre polynomial $P_{l}^{m}\left(\cos\theta\right)$
as \cite{key-43},

\begin{equation}
\ \ \ \ \ \ \ \ \ \ \ \ \ _{j}\mathcal{P}_{l_{j}}^{l_{j-1}}\left(\theta_{j}\right)=\sqrt{\frac{2l_{j}+j-1}{2}\frac{\left(l_{j}+l_{j-1}+j-2\right)!}{\left(l_{j}-l_{j-1}\right)!}}\sin^{\frac{2-j}{2}}\theta_{j}P_{l_{j}+\frac{j-2}{2}}^{-\left(l_{j-1}+\frac{j-2}{2}\right)}\left(\cos\theta_{j}\right).\label{eq:11}
\end{equation}
Thus the spherical harmonic function $\mathcal{Y}\left(\theta_{n},\theta_{n-1},...\theta_{1}\right)$
satisfies the eigen equation,

\begin{equation}
\nabla_{S^{n}}^{2}\mathcal{Y}_{l_{n},l_{n-1},...l_{1}}\left(\theta_{n},\theta_{n-1},...\theta_{1}\right)=-l_{n}\left(l_{n}+n-1\right)\mathcal{Y}_{l_{n},l_{n-1},...l_{1}}\left(\theta_{n},\theta_{n-1},...\theta_{1}\right),
\end{equation}
and the normalization condition

\begin{equation}
\int\mathcal{Y}_{l_{n}^{\prime},l_{n-1}^{\prime},...l_{1}^{\prime}}^{*}\mathcal{Y}_{l_{n},l_{n-1},...l_{1}}dV_{S^{n}}=\prod_{j=1}^{n}\delta_{l_{j}^{\prime}l_{j}}.
\end{equation}
$\chi_{n_{Z}}\left(Z\right)$ represents the eigenfunction of $H_{Z}$
which is obviously the eigenfunction of one-dimensional harmonic oscillator.
Keeping these in hand, the eigenvalue of Hamiltonian (\ref{eq:3})
is obtained as,

\begin{equation}
E_{n_{Z},n_{\rho},l_{n}}\left(N_{c},n\right)=E_{n_{Z}}+E_{n_{\rho},l_{n}}\left(N_{c},n\right)+M_{0},\label{eq:14}
\end{equation}
where

\begin{equation}
E_{n_{Z}}=\left(n_{Z}+\frac{1}{2}\right)\omega_{Z},\ E_{n_{\rho},l_{n}}\left(N_{c},n\right)=\left[\sqrt{2K+\left(l_{n}+\frac{n-1}{2}\right)^{2}}+2n_{\rho}+1\right]\omega_{\rho}.\label{eq:15}
\end{equation}

\subsection{The formula of QTOC in quantum mechanics}

In quantum mechanics, \cite{key-3,key-4,key-5} suggest the thermal
canonical and microcanonical OTOC $C_{T},c_{n}$ can be defined respectively
as,

\begin{equation}
C_{T}\left(t\right)=\frac{1}{\mathcal{Z}}\sum_{n}e^{-\beta E_{n}}c_{n}\left(t\right),\ \mathcal{Z}=\sum_{n}e^{-\beta E_{n}},\ c_{n}\left(t\right)=-\left\langle n\left|\left[x\left(t\right),p\right]^{2}\right|n\right\rangle ,\label{eq:16}
\end{equation}
for a time-independent Hamiltonian $H\left(x_{1},x_{2}...x_{J},p_{1},p_{2}...p_{J}\right)$.
Here $\mathcal{Z}$ is the partition function, $\beta=1/T$ refers
to the temperature of the system, $\left|n\right\rangle $ refers
to the $n$-th eigenstate of Hamiltonian satisfying the eigen equation
$H\left|n\right\rangle =E_{n}\left|n\right\rangle $ and $x\left(t\right),p\left(t\right)$
is respectively the canonical coordinate and momentum in Heisenberg
picture. Let us denote one of the canonical coordinates and momentums
as,

\begin{equation}
x\left(t\right)=x_{a}\left(t\right),p\left(t\right)=p_{a}\left(t\right),p=p\left(0\right),a\in\left\{ 1,2...J\right\} ,
\end{equation}
for notational simplicity. Using the completeness condition in quantum
mechanics, the OTOC can be written as,

\begin{equation}
c_{n}\left(t\right)=-\sum_{m}\left\langle n\left|\left[x\left(t\right),p\right]\right|m\right\rangle \left\langle m\left|\left[x\left(t\right),p\right]\right|n\right\rangle \equiv\sum_{m}b_{nm}\left(t\right)b_{nm}^{*}\left(t\right),\label{eq:18}
\end{equation}
where

\begin{equation}
b_{nm}\left(t\right)=-i\left\langle n\left|\left[x\left(t\right),p\right]\right|m\right\rangle .\label{eq:19}
\end{equation}
Recall the unitary transformation $x\left(t\right)=e^{iHt}xe^{-iHt}$,
(\ref{eq:19}) can be further rewritten as,

\begin{equation}
b_{nm}\left(t\right)=-i\sum_{k}\left(e^{iE_{nk}t}x_{nk}p_{km}-e^{iE_{km}t}p_{nk}x_{km}\right),
\end{equation}
where $x_{nk}=\left\langle n\left|x\right|k\right\rangle ,p_{nk}=\left\langle n\left|p\right|k\right\rangle $
and $E_{nk}=E_{n}-E_{k}$. If a time-independent Hamiltonian is given
as,

\begin{equation}
H=\sum_{a=1}^{J}\frac{p_{a}^{2}}{2m_{X}}+U\left(x_{1},x_{2}...x_{J}\right),
\end{equation}
the following relation can be obtained as,

\begin{equation}
p_{mn}=im_{X}E_{mn}x_{mn},\label{eq:20}
\end{equation}
by the commutation relation $\left[H,x\right]=-\frac{ip}{m_{X}}$
which leads to 

\begin{equation}
b_{nm}\left(t\right)=m_{X}\sum_{k}x_{nk}x_{km}\left(E_{km}e^{iE_{nk}t}-E_{nk}e^{iE_{km}t}\right).\label{eq:23}
\end{equation}
Therefore the OTOCs can be obtained by using (\ref{eq:16}) (\ref{eq:18})
and (\ref{eq:23}). We note that since $H_{Z}$ presented in (\ref{eq:3})
is nothing but the Hamiltonian of one-dimensional harmonic oscillator,
it OTOC can be analytically obtained as \cite{key-3,key-4},

\begin{equation}
C_{T}\left(t\right)=c_{n_{Z}}\left(t\right)=m_{Z}^{2}\cos^{2}\omega_{Z}t.\label{eq:24}
\end{equation}
Accordingly, in the following discussion, we will focus on the numerical
evaluation of the OTOCs with respect to $H_{y}$. 

\section{The numerical analysis}

\begin{figure}
\begin{centering}
\includegraphics[scale=0.37]{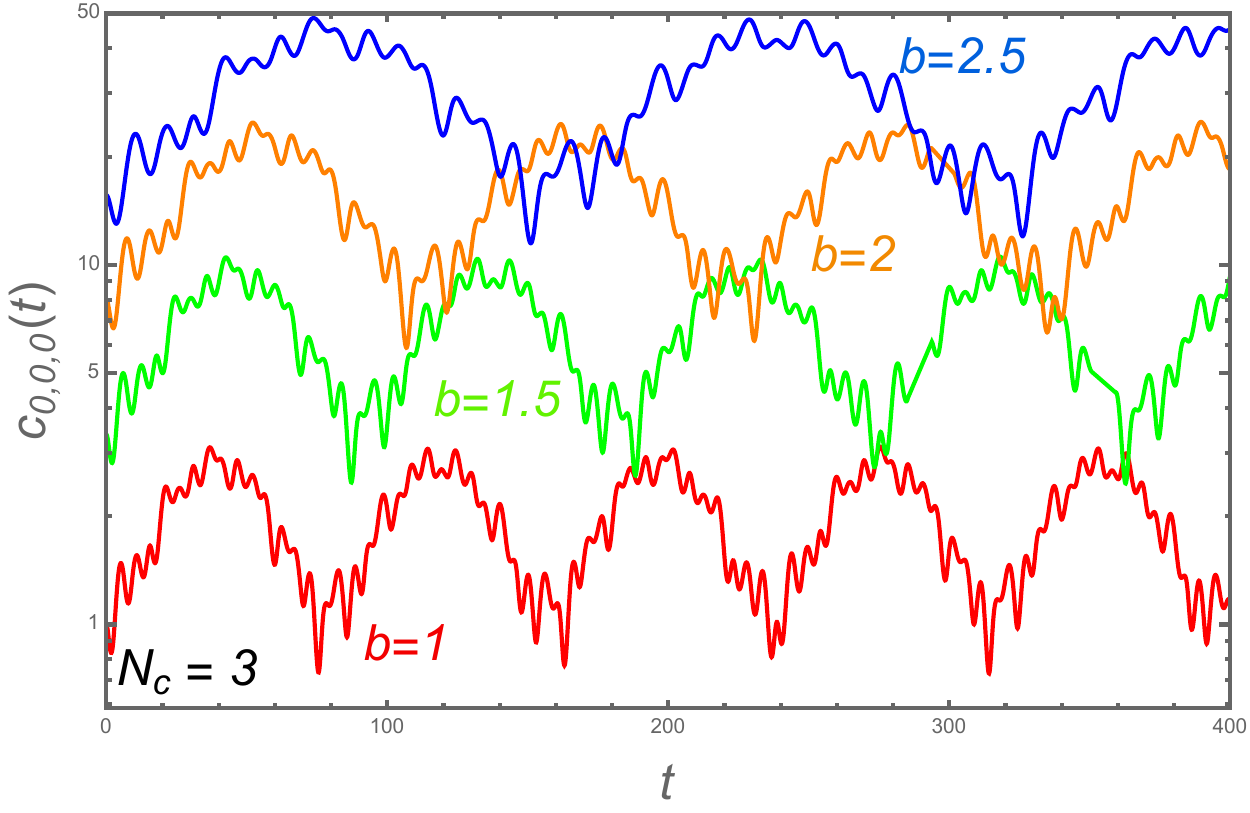}\includegraphics[scale=0.37]{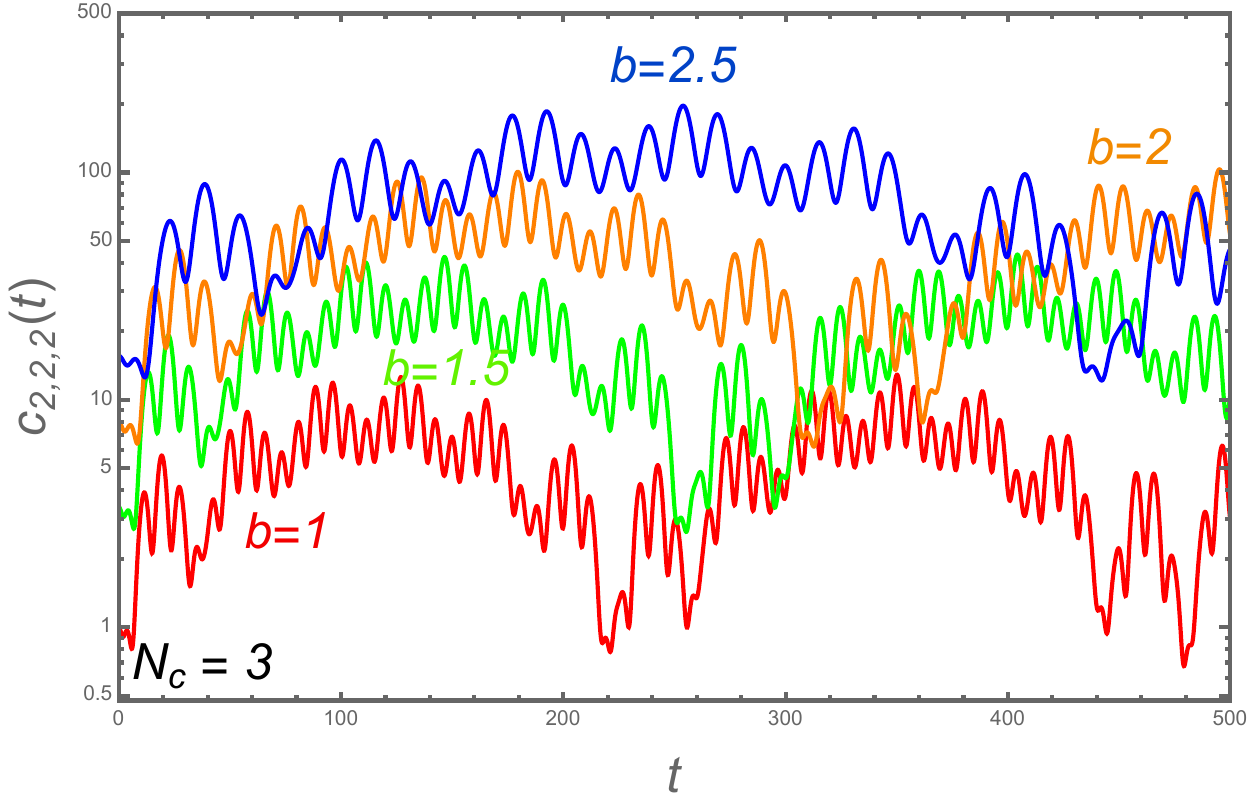}
\par\end{centering}
\begin{centering}
\includegraphics[scale=0.37]{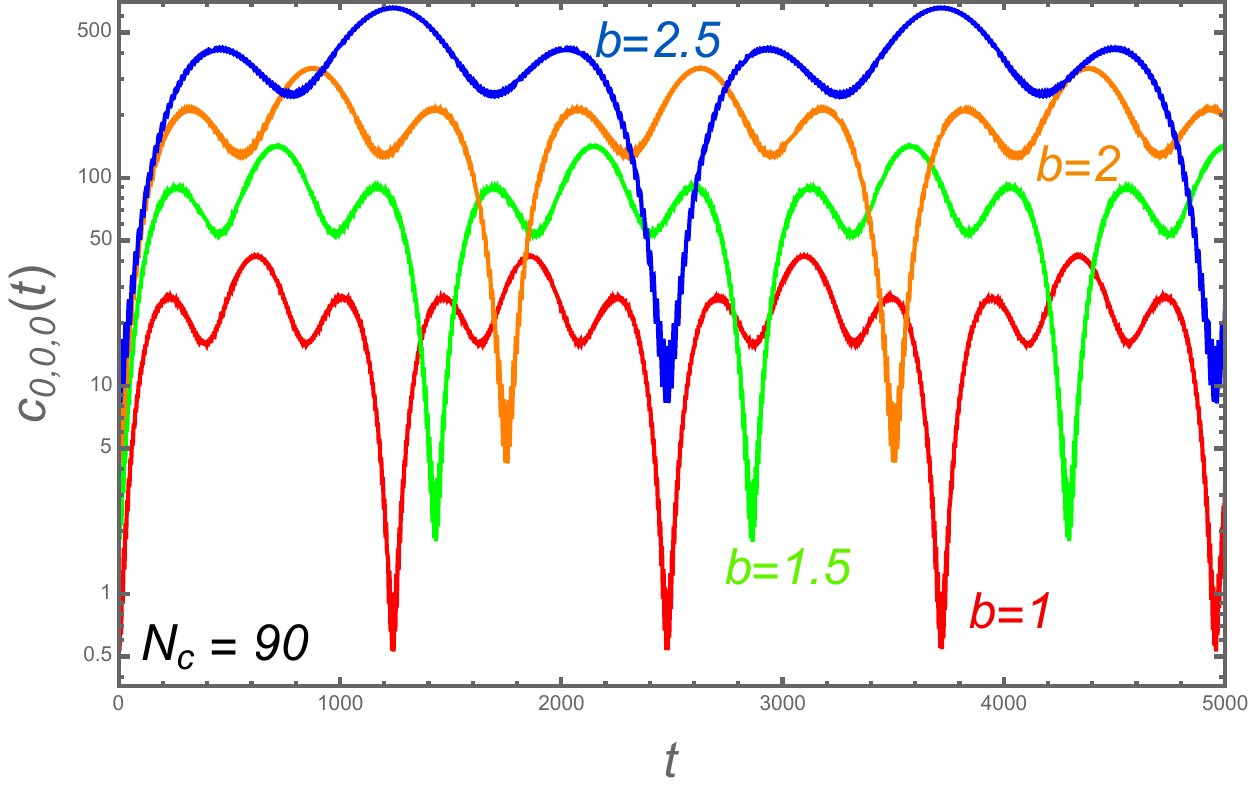}\includegraphics[scale=0.37]{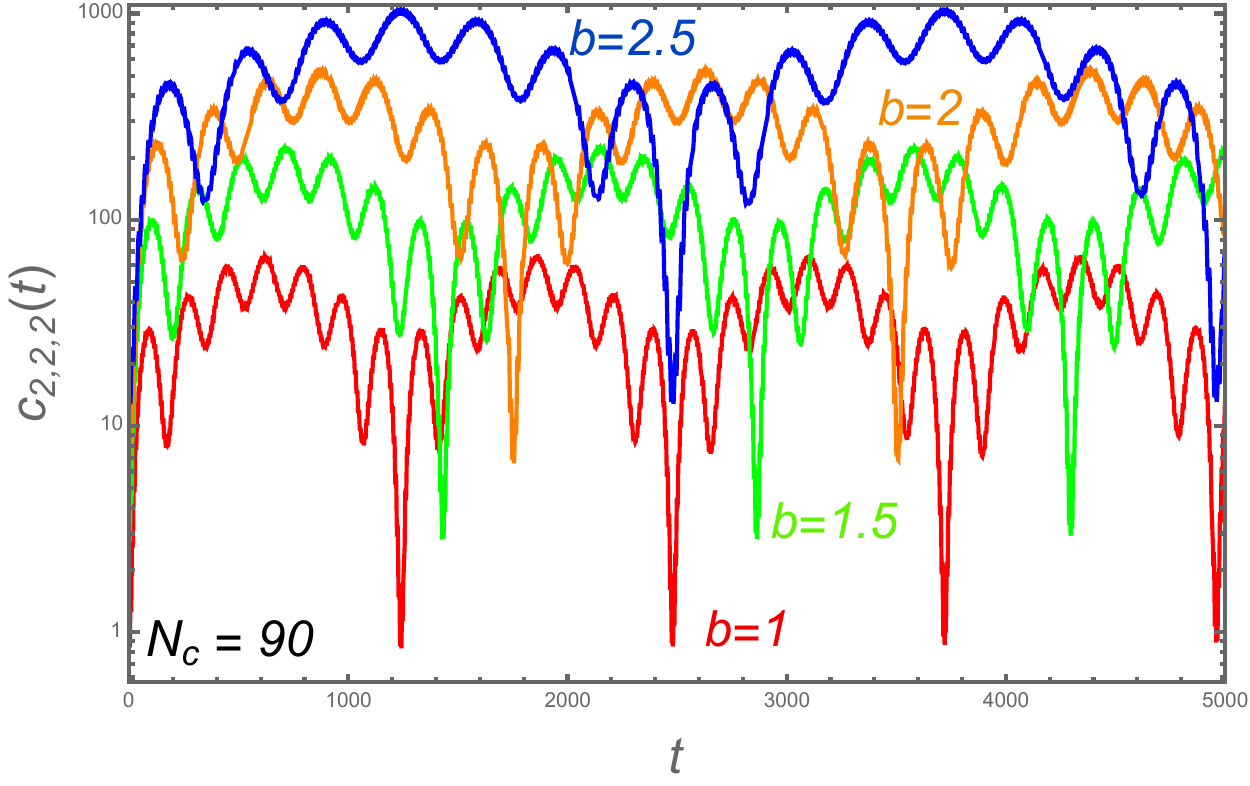}
\par\end{centering}
\caption{\label{fig:1} The two-flavored ($N_{f}=2$) microcanonical OTOCs
$c_{n_{\rho},l_{3},l_{2}}$ ($c_{0,0,0},c_{2,2,2}$) as functions
of time $t$ with various $b$.}

\end{figure}
 
\begin{figure}[t]
\begin{centering}
\includegraphics[scale=0.37]{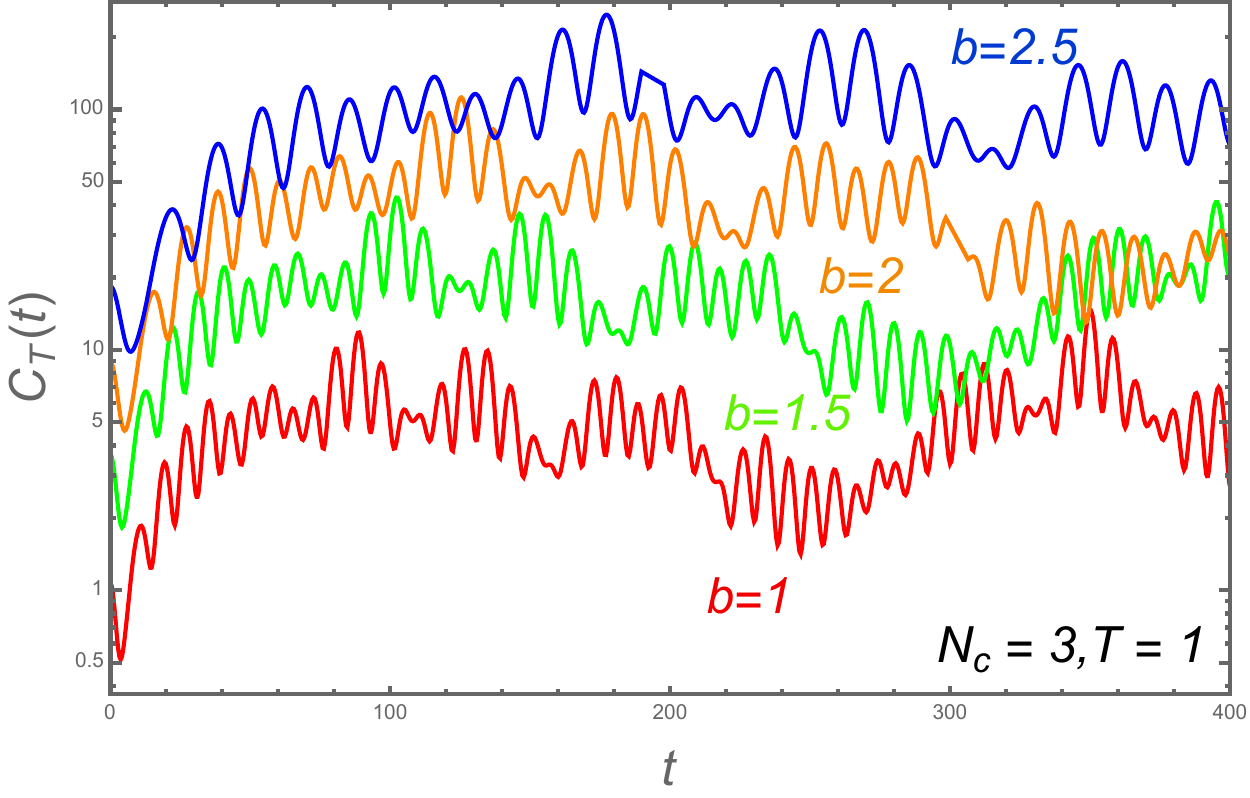}\includegraphics[scale=0.37]{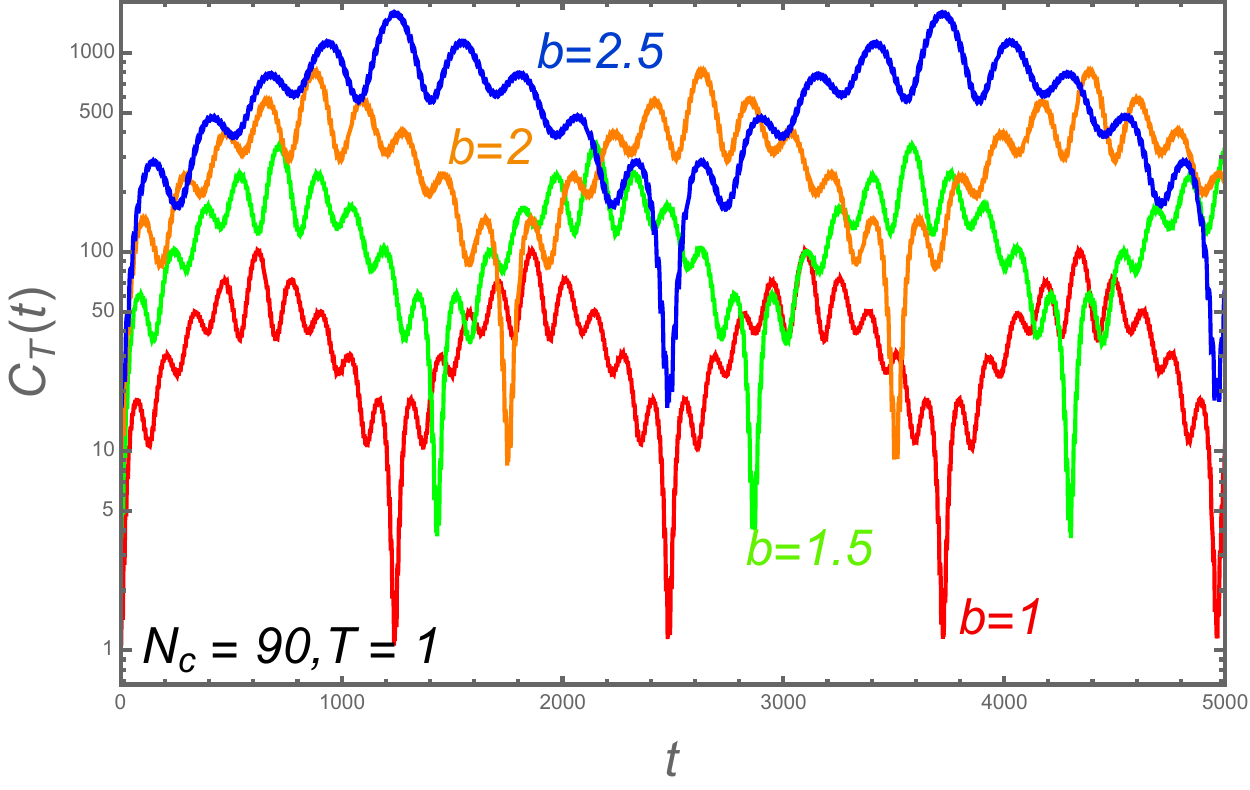}
\par\end{centering}
\begin{centering}
\includegraphics[scale=0.37]{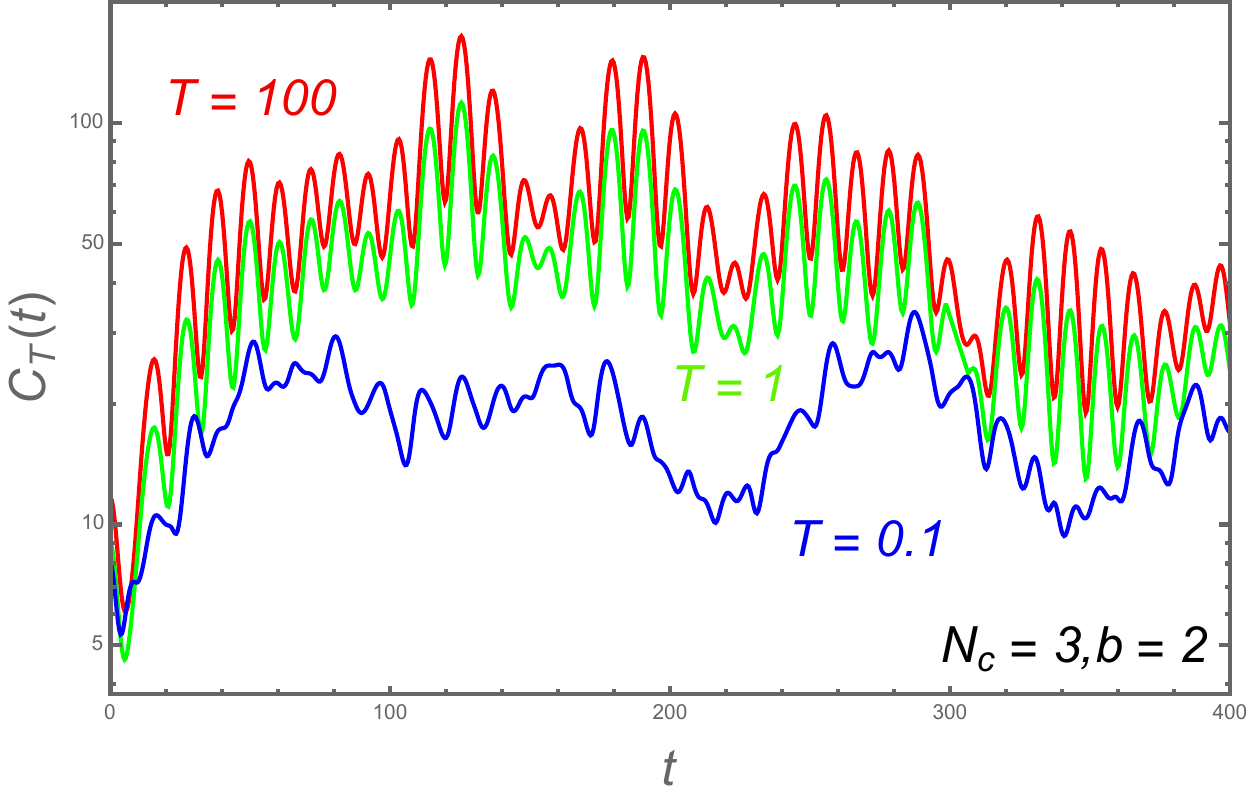}\includegraphics[scale=0.37]{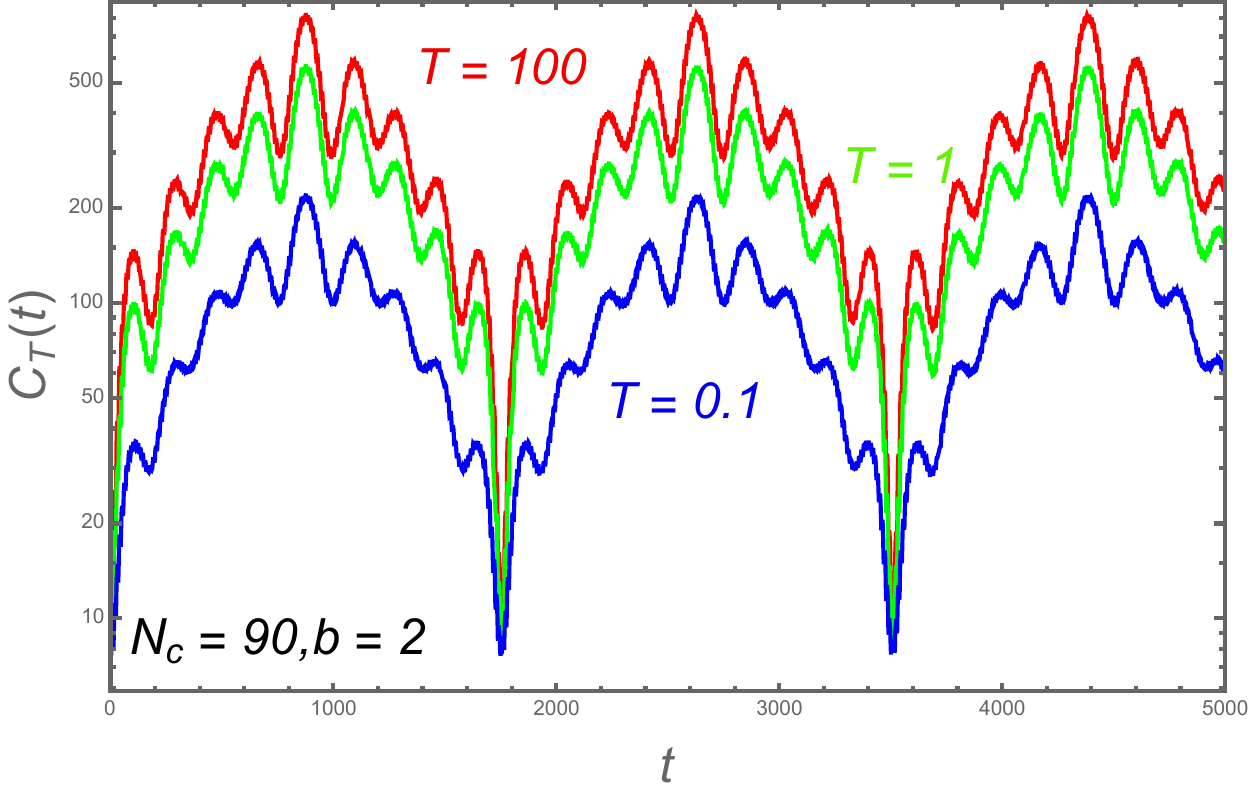}
\par\end{centering}
\caption{\label{fig:2} The two-flavored ($N_{f}=2$) thermal OTOCs $C_{T}$
as functions of time $t$ with various $b$ and $T$.}
\end{figure}
 
\begin{figure}[t]
\begin{centering}
\includegraphics[scale=0.36]{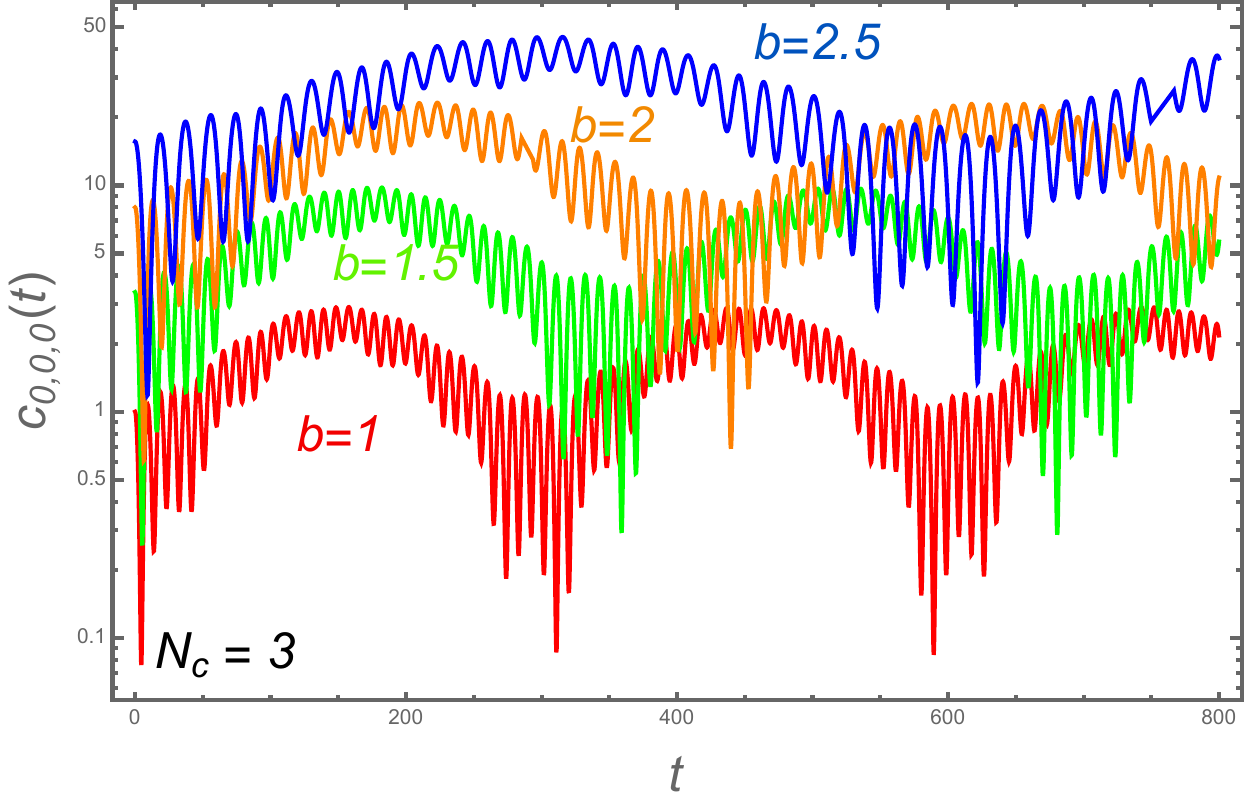}\includegraphics[scale=0.36]{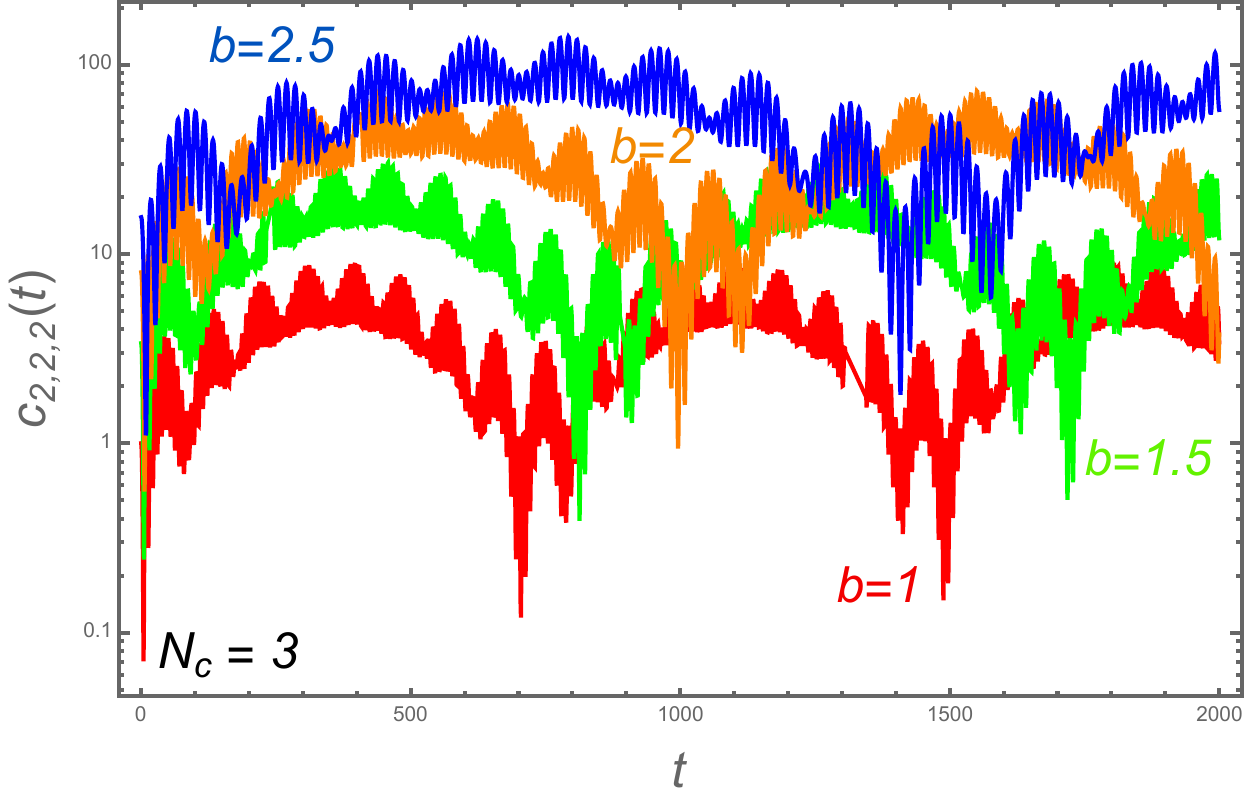}
\par\end{centering}
\begin{centering}
\includegraphics[scale=0.36]{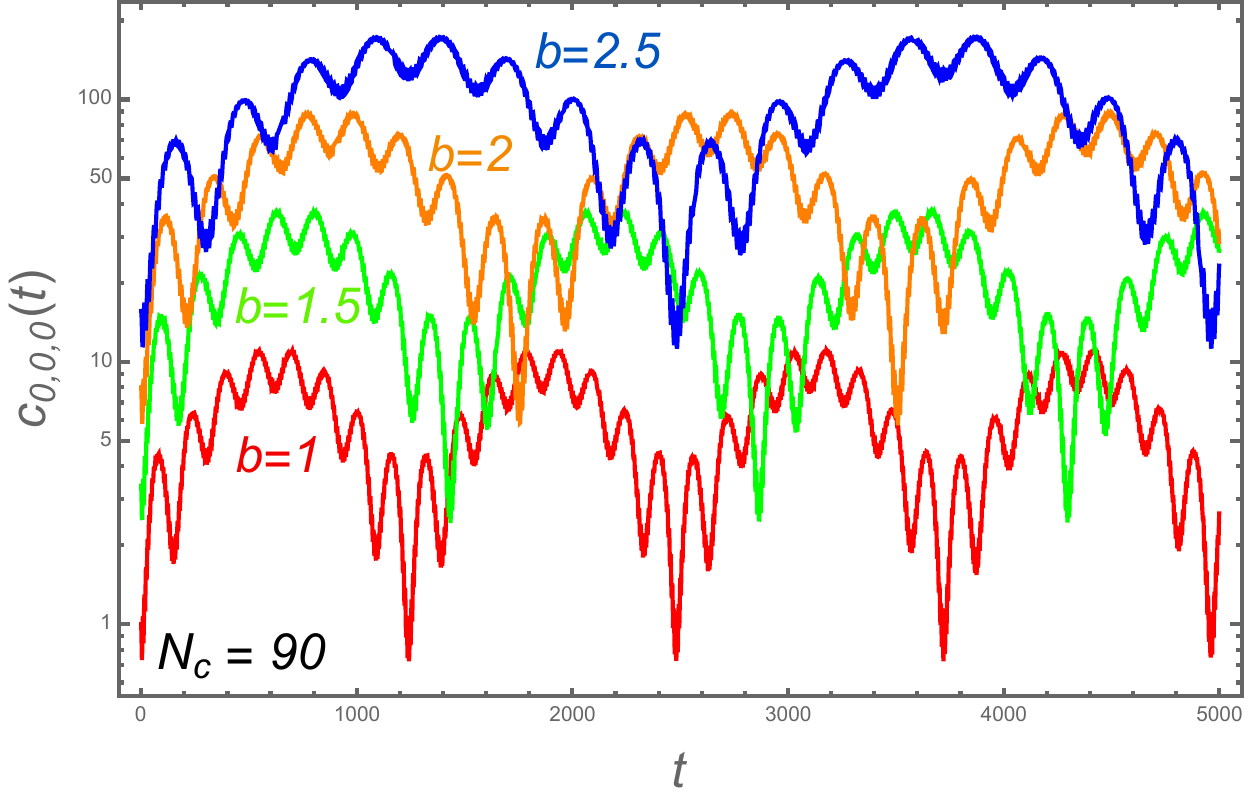}\includegraphics[scale=0.36]{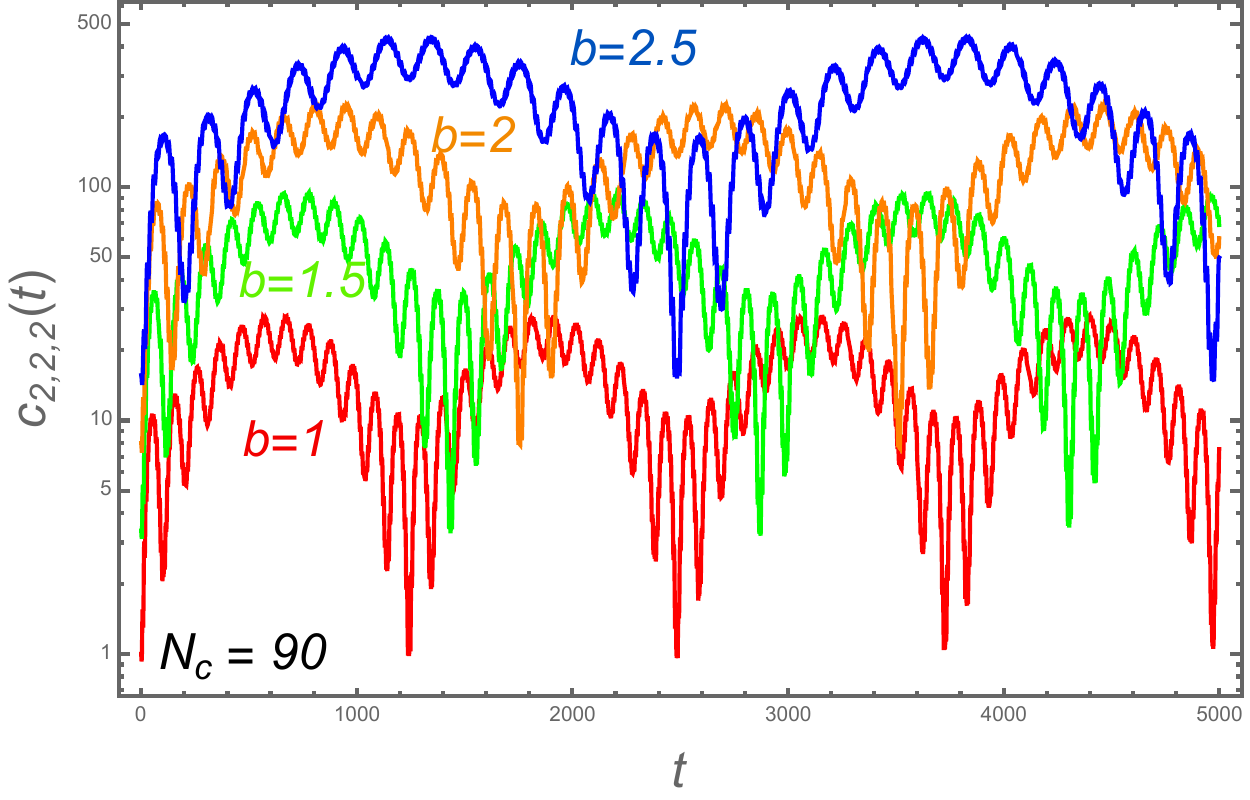}
\par\end{centering}
\caption{\label{fig:3} The three-flavored ($N_{f}=3$) microcanonical OTOCs
$c_{n_{\rho},l_{8},l_{7}}$ ($c_{0,0,0},c_{2,2,2}$) as functions
of time $t$ with various $b$.}
\end{figure}
 
\begin{figure}[t]
\begin{centering}
\includegraphics[scale=0.36]{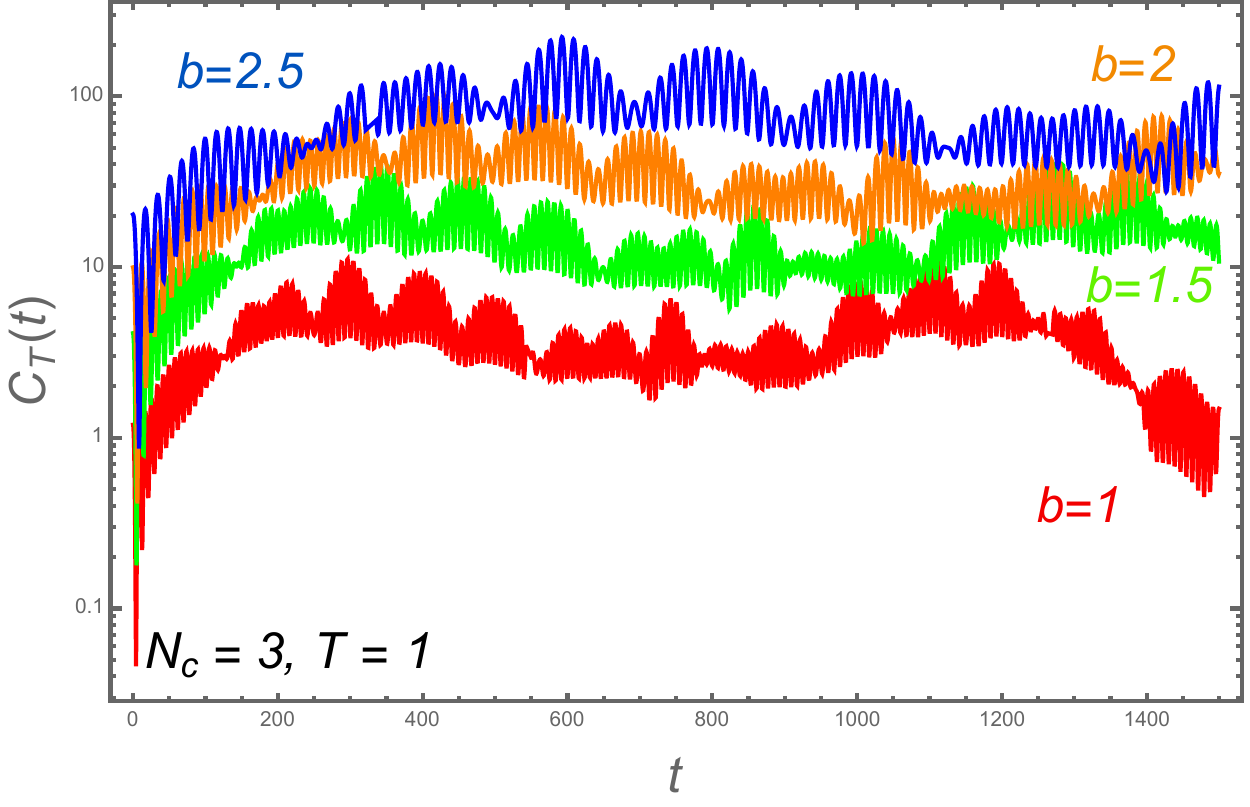}\includegraphics[scale=0.36]{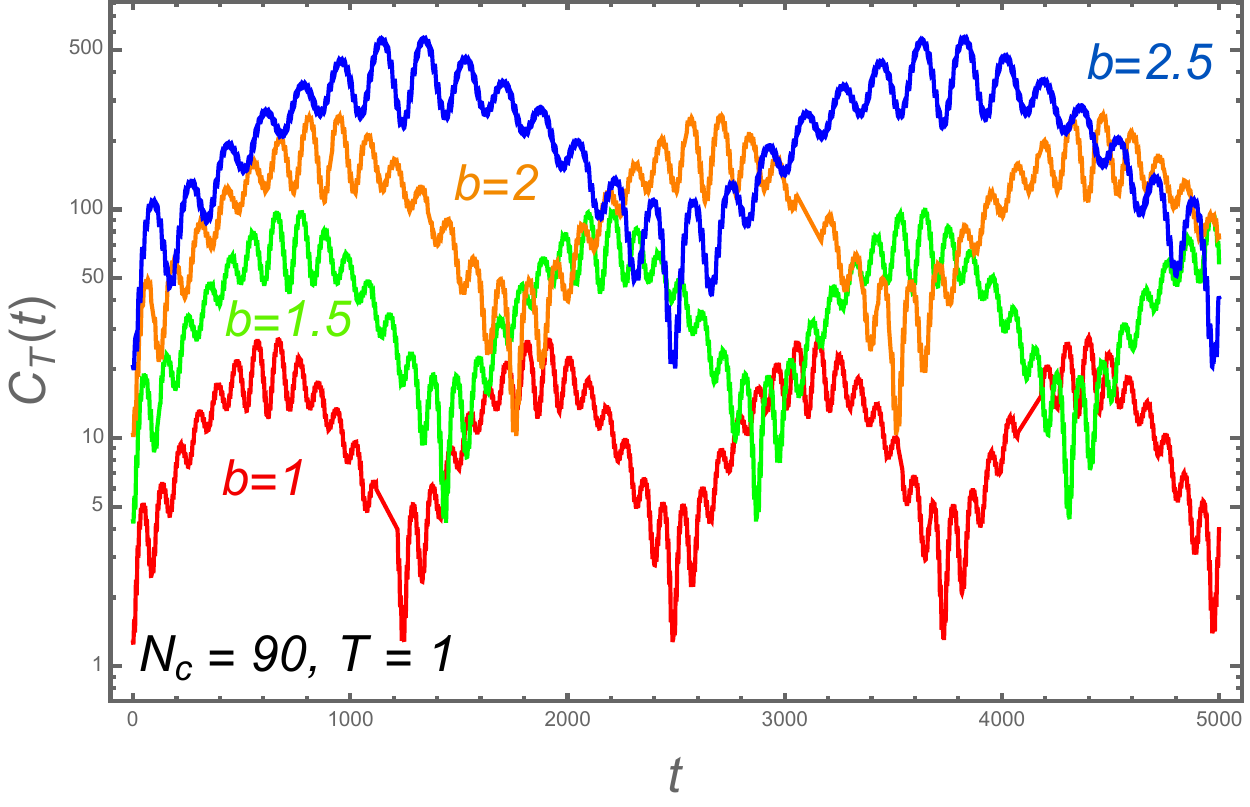}
\par\end{centering}
\begin{centering}
\includegraphics[scale=0.37]{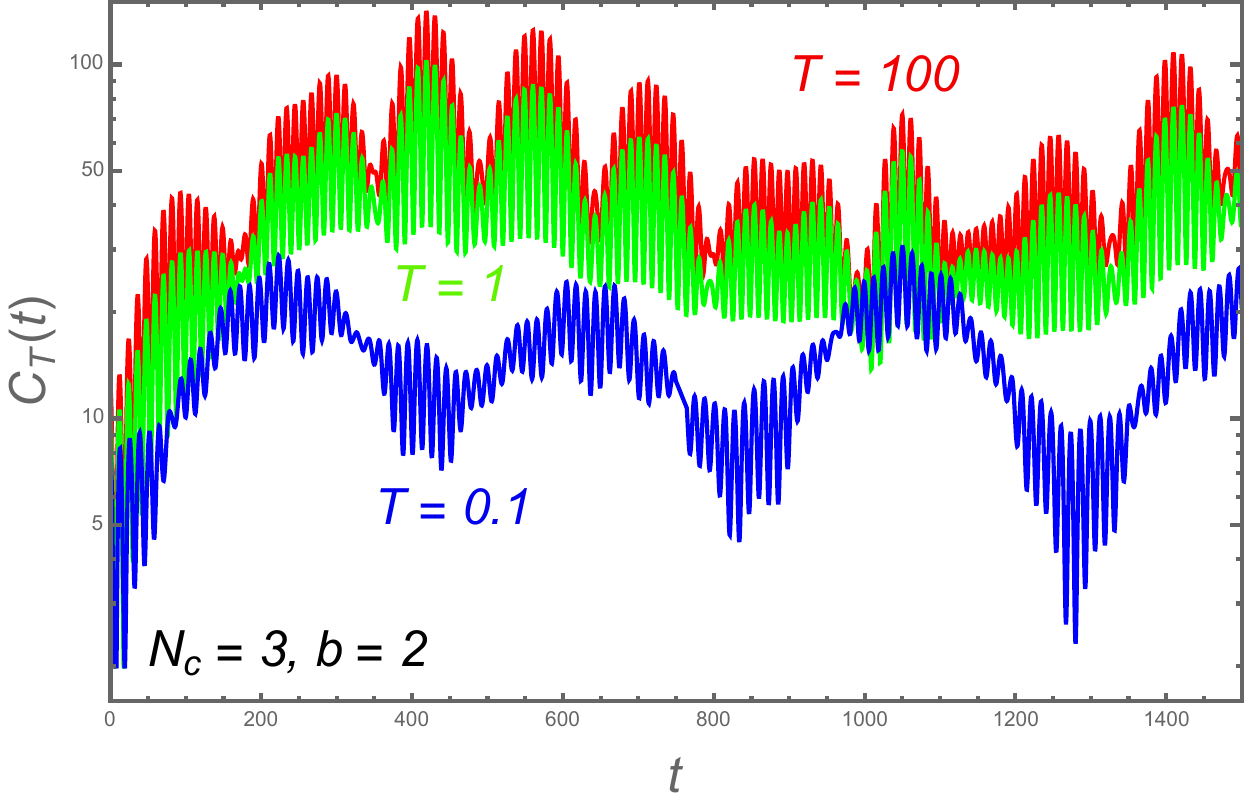}\includegraphics[scale=0.37]{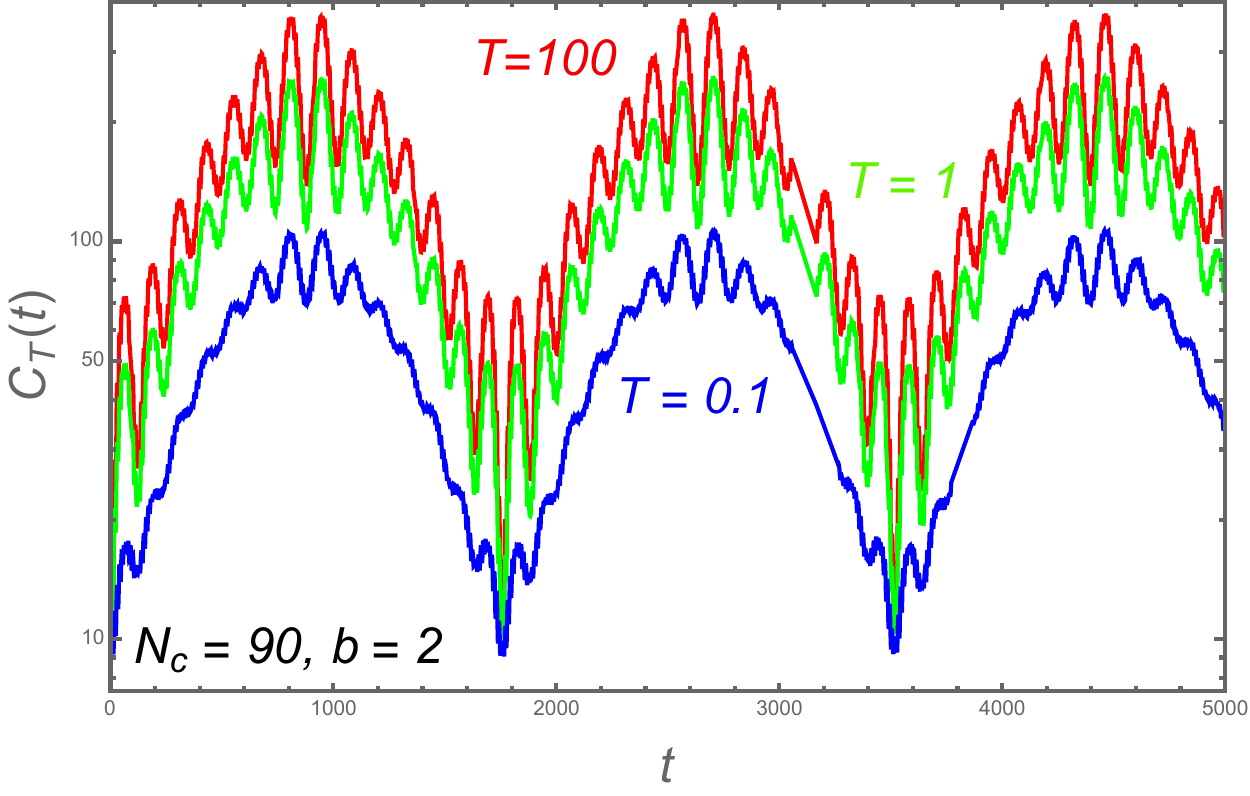}
\par\end{centering}
\caption{\label{fig:4} The three-flavored ($N_{f}=3$) thermal OTOCs $C_{T}$
as functions of time $t$ with various $b$ and $T$.}
\end{figure}
 
\begin{figure}[t]
\begin{centering}
\includegraphics[scale=0.37]{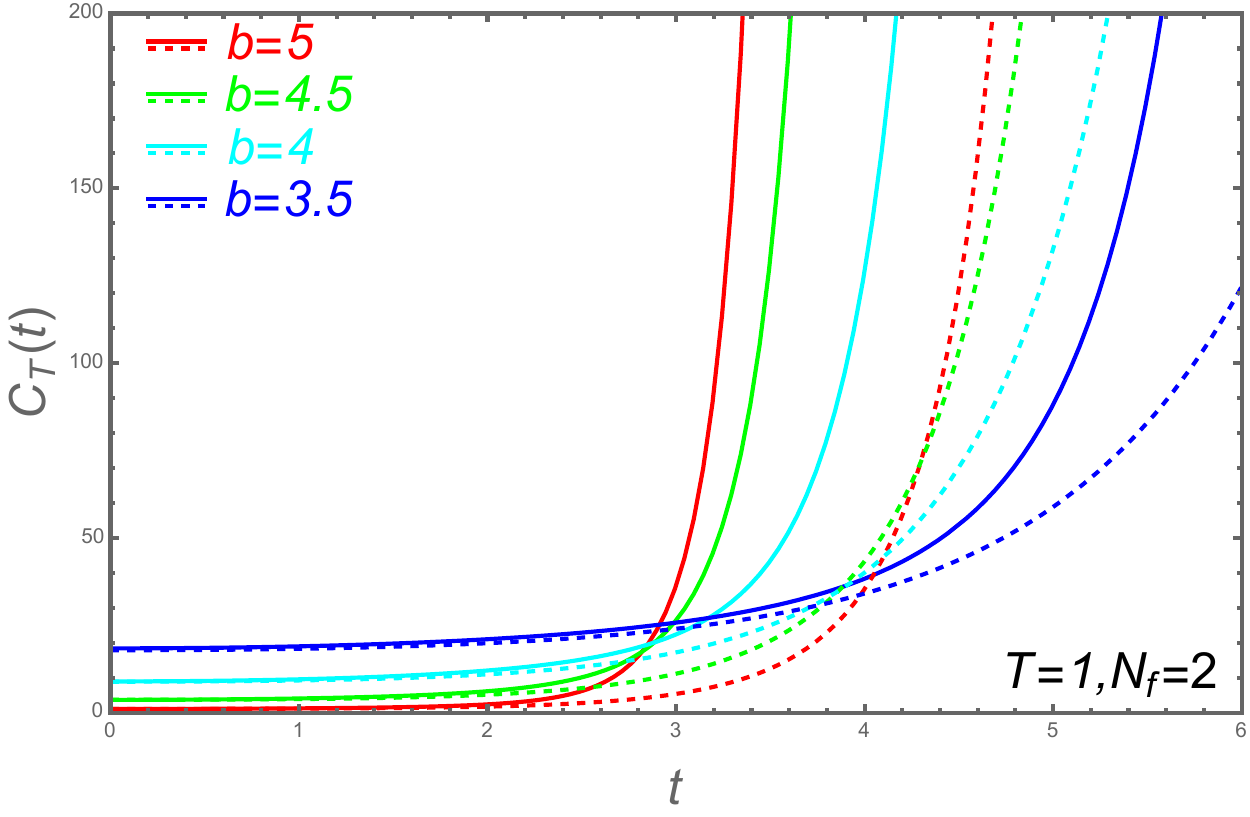}\includegraphics[scale=0.37]{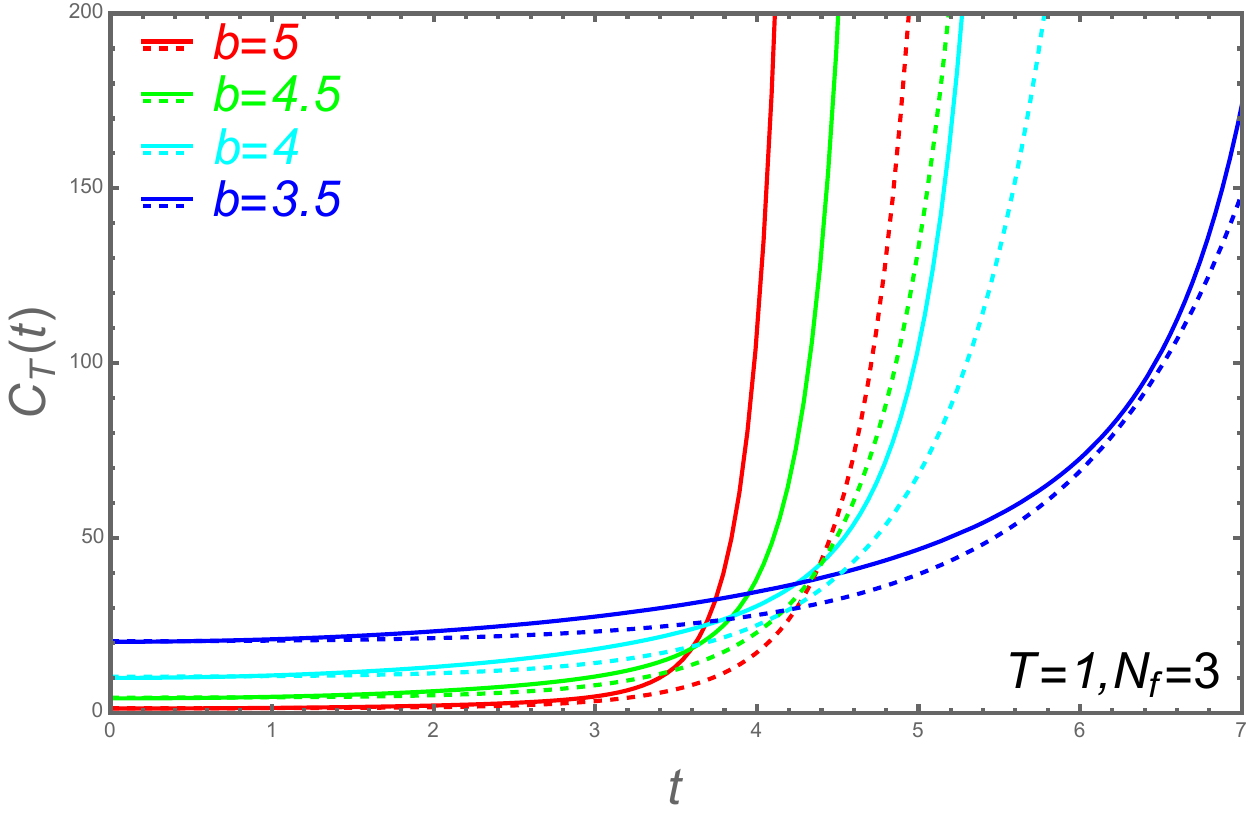}
\par\end{centering}
\begin{centering}
\includegraphics[scale=0.37]{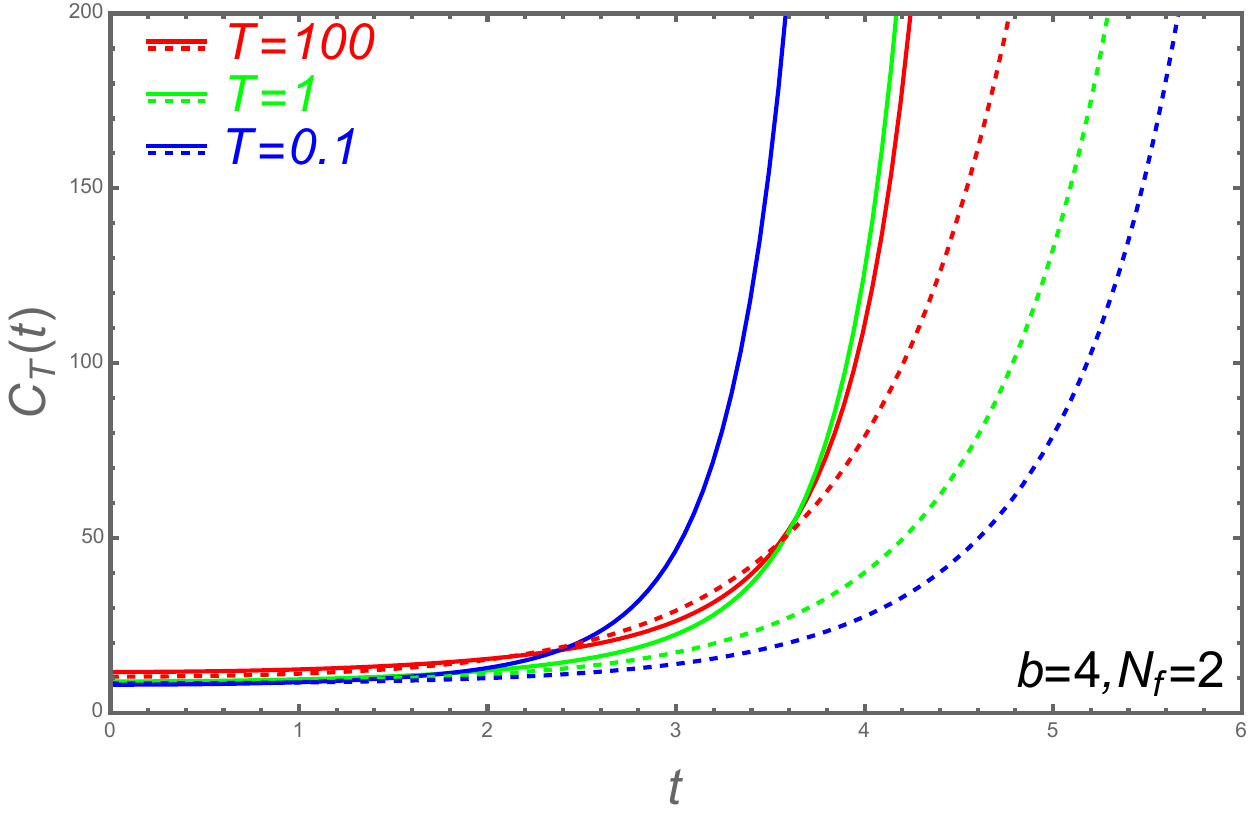}\includegraphics[scale=0.37]{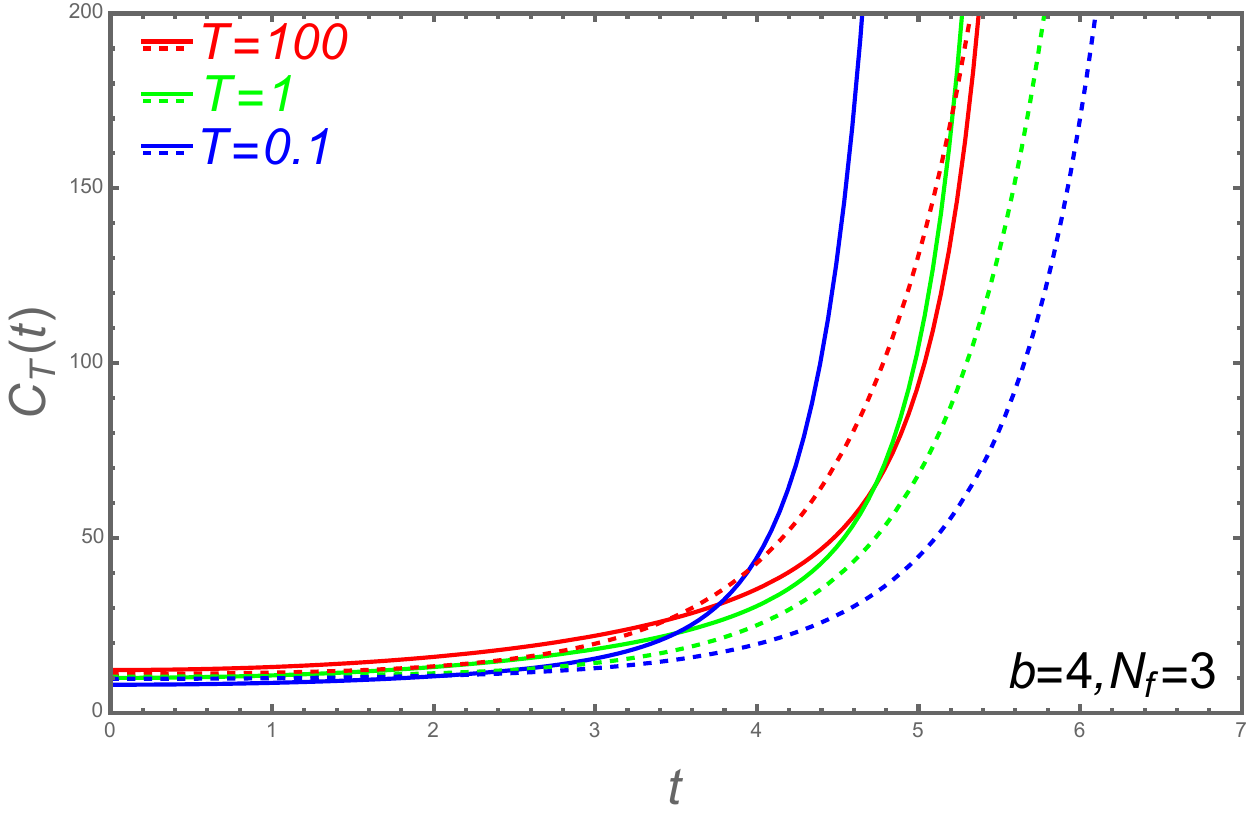}
\par\end{centering}
\caption{\label{fig:5}The thermal OTOCs $C_{T}$ as functions of time $t$
for $b>3$. All the solid lines correspond to $N_{c}=3$ while all
the dashed lines correspond to $N_{c}=90$.}

\end{figure}
In this section, we focus on the OTOC with respect to $H_{y}$ presented
in (\ref{eq:3}). The microcanonical and thermal OTOCs $c_{n_{\rho},l_{n},l_{n-1}...l_{1}}\left(t\right),C_{T}\left(t\right)$
of the holographic Skyrmion with various $b$ are given in Figure
\ref{fig:1} - \ref{fig:4}. We note that the quantum numbers $l_{n-2},l_{n-3},...l_{2},l_{1}$
are degenerated quantum numbers, thus we denote $c_{n_{\rho},l_{n},l_{n-1}...l_{1}}\left(t\right)$
as $c_{n_{\rho},l_{n},l_{n-1}}\left(t\right)$. The sum presented
in (\ref{eq:18}) and (\ref{eq:23}) has been truncated at $n_{c}=4$
i.e. $0\leq n_{Z},n_{\rho},l_{n}\leq n_{c}$ for the actual numerical
calculation and the degeneracy number $n_{\mathrm{d}}$ in $c_{n_{\rho},l_{n},l_{n-1}...l_{1}}\left(t\right)$
is computed as,

\begin{equation}
n_{\mathrm{d}}\left(l_{n}\right)=\sum_{l_{n-1}=0}^{l_{n}}\sum_{l_{n-2}=0}^{l_{n-1}}...\sum_{l_{3}=0}^{l_{4}}\sum_{l_{2}=0}^{l_{3}}\left(2l_{2}+1\right).\label{eq:25}
\end{equation}
Thus it includes totally $\left(n_{c}+1\right)^{2}\times\sum_{l_{3}=0}^{n_{c}}n_{\mathrm{d}}\left(l_{3}\right)=1375$
states for $N_{f}=2$ and $\left(n_{c}+1\right)^{2}\times\sum_{l_{8}=0}^{n_{c}}n_{\mathrm{d}}\left(l_{8}\right)=16500$
states for $N_{f}=3$ contributing to the OTOC. Our numerical calculation
reveals the following features of the OTOC. First, the thermal OTOCs
do not display the apparent exponential growth while they grow at
early times. Thus the exponential growth expected in the classical
OTOC is not apparent. Second, the microcanonical OTOCs are approximately
periodic because of the commensurable energy spectrum. For the situation
$N_{c}\gg1$, the period $\Delta t$ of microcanonical OTOCs is evaluated
as $\Delta t\sim\mathrm{min}\left[E_{n_{\rho},l_{n};n_{\rho}^{\prime},l_{n}^{\prime}}^{-1}\right]\sim N_{c}$
for $N_{c}\rightarrow\infty$ which displays the large-$N_{c}$ behaviors
of the microcanonical OTOCs as \cite{key-4}. And it does not depend
on temperature basically. Third, the thermal OTOCs grow with $T$
and $b$ (the presence of instanton or theta angle) according to (\ref{eq:5})
and in addition they tend to become periodic at large $N_{c}$ as
the microcanonical OTOCs. 

Keep these in hand, let us attempt to outline how to use OTOC of the
holographic Skyrmion to describe the various status of the baryonic
phase with instanton charge denoted as $b$ in this model. As the
OTOC is periodic for large $N_{c}$, it means the quantum mechanical
system (\ref{eq:3}) is not really chaotic. Nevertheless, we can define
imaginary Lyapunov coefficient as $\tilde{L}=iL$ (where $L$ is real)
as it is suggested in \cite{key-4} to characterize the periodic quantum
OTOC $C_{T}$ as $C_{T}\left(t\right)\sim e^{\tilde{L}t}$. Therefore,
according to the numerical calculation presented in Figure \ref{fig:1}
- \ref{fig:4}, we can see that the Lyapunov coefficient $L$ decreases
with the growth of $b$ which is a function as $L\left(b\right)$.
On the other hand, since the excitations of holographic Skyrmion in
the D0-D4/D8 system is recognized as baryon states \cite{key-27,key-28,key-29,key-40,key-41},
it means $L\left(b\right)$ as an order parameter indicates possibly
baryonic states with various $b$ or theta angle. Recall the holographic
description of the metastable baryonic states in \cite{key-28,key-29},
it seemingly implies the value of $L$ with $1\leq b<3$ represents
the metastable baryonic states and the maximal value of $L\left(b=1\right)$
indicates the baryonic states become stable. So the Lyapunov coefficient
as a function of instanton charge indicates the metastability or stability
of the baryonic phase.

Moreover, although the derivation with Hamiltonian for Skyrmion (\ref{eq:3})
is strictly valid only for $1\leq b\leq3$, it would be interesting
to extend our discussion with $b>3$. According to (\ref{eq:4}) (\ref{eq:14})
and (\ref{eq:23}), the energy difference $E_{nk}$ presented in the
exponents becomes totally imaginary for $b>3$ for the part described
by $H_{y}$, so the OTOC will contain the factor $e^{\left|E_{nk}\right|t}$
i.e. the real Lyapunov exponent due to the derivations in Section
2.2. Therefore we could simply replace $E_{nk}\rightarrow iE_{nk}$
in the presented exponents then compute the OTOC with the same derivation
as the case of $1<b<3$. The corresponding numerical calculations
are illustrated in Figure \ref{fig:5}. So we can see the OTOCs become
really exponential as $e^{Lt}$ which implies the system for $b>3$
is really chaotic. And the Lyapunov coefficient also trends to be
vanished at large $N_{c}$ which is consistent with that the dual
gravity bulk does not have a horizon in this model \cite{key-19,key-20,key-21,key-22}.
This behavior of OTOC with $b>3$ could be more explicit by recalling
its $Z$ part given in (\ref{eq:24}). For $b>3$, $\omega_{Z}$ is
purely imaginary so that we have $C_{T}\sim\cos^{2}\omega_{Z}t\rightarrow e^{2\left|\omega_{Z}\right|t}$
at large $t$ which illustrates indeed the system becomes really chaotic.
This is reasonable since the harmonic oscillator with imaginary frequency
is totally unstable and very sensitive to its initial state. Altogether,
we can see the imaginary Lyapunov exponent indicates the metastability
or stability of the baryonic phase while the Lyapunov exponent becomes
real indicating its instability. In this sense, we may conclude that
the Lyapunov coefficient could be treated as an order parameter to
indicate the baryonic phase structure in our holographic model. 

\section{The classical limit of the OTOC}

According to the definition (\ref{eq:16}), the classical version
of OTOC can be naturally obtained by replacing the quantum commutators
with Poisson bracket, i.e. $\frac{1}{i}\left[,\right]\rightarrow\left\{ ,\right\} _{\mathrm{P.B}}$.
And thermal average must be replaced by the integral in the phase
space consisted of canonical coordinate $\mathbf{q}$ and momentum
$\mathbf{p}$. Therefore, we can obtain the classical version of the
OTOC presented in (\ref{eq:16}) as,

\begin{equation}
C_{T}^{\mathrm{cl}}\left(t\right)=\frac{1}{\mathcal{Z}_{\mathrm{cl}}}\int\frac{d\mathbf{q}d\mathbf{p}}{\left(2\pi\right)^{n+2}}e^{-\beta H}\left\{ \mathbf{q}\left(t\right),\mathbf{p}\left(0\right)\right\} _{\mathrm{P.B}}^{2},c^{\mathrm{cl}}\left(t\right)=\left\{ \mathbf{q}\left(t\right),\mathbf{p}\left(0\right)\right\} _{\mathrm{P.B}}^{2},
\end{equation}
where

\begin{equation}
\mathcal{Z}_{\mathrm{cl}}=\int\frac{d\mathbf{q}d\mathbf{p}}{\left(2\pi\right)^{n+2}}e^{-\beta H}.
\end{equation}
Besides, the classical version of the Hamiltonian (\ref{eq:3}) of
the Skyrmion is given by

\begin{align}
H= & M_{0}+H_{Z}+H_{y},\nonumber \\
H_{Z}= & \frac{p_{Z}^{2}}{2m_{Z}}+\frac{1}{2}m_{Z}\omega_{Z}^{2}Z^{2},\nonumber \\
H_{y}= & \frac{1}{2m_{\rho}}\sum_{I=1}^{n+1}p_{I}^{2}+\frac{1}{2}m_{\rho}\omega_{\rho}^{2}\rho^{2}+\frac{K}{m_{\rho}\rho^{2}}.\label{eq:28}
\end{align}
As the canonical coordinate $\mathbf{q}$ and momentum $\mathbf{p}$
is chosen as $\mathbf{q}=\left\{ Z,y_{I}\right\} ,\mathbf{p}=\left\{ p_{Z},p_{I}\right\} $,
for the $Z$ part in Hamiltonian (\ref{eq:28}), its classical solution
is given as,

\begin{align}
Z\left(t\right) & =Z\cos\omega_{Z}t+\frac{p_{Z}}{m_{Z}}\sin\omega_{Z}t,\nonumber \\
p_{Z}\left(t\right) & =-m_{Z}\omega_{Z}\sin\omega_{Z}t+p_{Z}\omega_{Z}\cos\omega_{Z}t.\label{eq:29}
\end{align}
Hence the associated OTOCs are computed as,

\begin{align}
c_{n_{Z}}^{\mathrm{cl}}\left(t\right) & =\left\{ Z\left(t\right),p_{Z}\left(0\right)\right\} _{\mathrm{P.B}}^{2}=\left[\frac{\delta Z\left(t\right)}{\delta Z\left(0\right)}\right]^{2}=\cos^{2}\omega_{Z}t,\nonumber \\
C_{T}^{\mathrm{cl}}\left(t\right) & =\frac{1}{\mathcal{Z}_{\mathrm{cl}}}\int\frac{d\mathbf{q}d\mathbf{p}}{\left(2\pi\right)^{n+2}}e^{-\beta H}\left\{ Z\left(t\right),p_{Z}\left(0\right)\right\} _{\mathrm{P.B}}^{2}=\cos^{2}\omega_{Z}t,
\end{align}
which agrees quantitatively with its quantum version. For the $y$
part in (\ref{eq:28}), the classical equation of motion is obtained
as,

\begin{equation}
m_{\rho}\ddot{y}_{I}=\left(-m_{\rho}\omega_{\rho}^{2}+\frac{2K}{m_{\rho}\rho^{4}}\right)y_{I}.\label{eq:31}
\end{equation}
Note that for the classical Skyrmion, $\rho$ refers to the size of
the Skyrmion (as instantons) which must minimize the classical Hamiltonian
(\ref{eq:28}), thus we have \cite{key-28}

\begin{equation}
\rho^{\mathrm{cl}}=\frac{\left(2K\right)^{1/4}}{\left(m_{\rho}\omega_{\rho}\right)^{1/2}}.\label{eq:32}
\end{equation}
Impose (\ref{eq:32}) into (\ref{eq:31}), the classical solution
for $y_{I}$ is

\begin{equation}
y_{I}\left(t\right)=y_{I}\left(0\right)+\frac{p_{I}\left(0\right)}{m_{\rho}}t,
\end{equation}
then the associated OTOCs are calculated as,

\begin{align}
c_{y_{I}}^{\mathrm{cl}} & =\left\{ y_{I}\left(t\right),p_{I}\left(0\right)\right\} _{\mathrm{P.B}}^{2}=1,\nonumber \\
C_{T}^{\mathrm{cl}}\left(t\right) & =\frac{1}{\mathcal{Z}_{\mathrm{cl}}}\int\frac{d\mathbf{q}d\mathbf{p}}{\left(2\pi\right)^{n+2}}e^{-\beta H}\left\{ y_{I}\left(t\right),p_{I}\left(0\right)\right\} _{\mathrm{P.B}}^{2}=1,
\end{align}
which does not display the apparent dependence on $b$ different from
the quantum OTOCs. In this sense, we comment that if the Lyapunov
coefficient could be an order parameter to characterize some properties
of QCD, it implies the instantonic or theta- dependence is basically
dominated by the quantum feature of QCD, as it is expected \cite{key-R1,key-R2,key-R3,key-36}.

\section{Summary }

In this work, we employ the definition of OTOC in quantum mechanics
to derive the formulas and demonstrate explicitly the numerical calculations
of the OTOC of the holographic Skyrmion in the D0-D4/D8 model which
is equivalent to QCD with instantons or a theta angle according to
the gauge-gravity duality. Our numerical calculation supports that
OTOC in integrable quantum mechanical system does not include exponential
growth thus such system is not really chaotic as it is discussed in
\cite{key-3,key-4,key-5}. In particular, we attempt to use the imaginary
Lyapunov coefficient as an order parameter to indicate the various
status of the baryonic phase which illustrates the possibly metastable
states of baryon with instantons as it is discussed in the existing
works \cite{key-21,key-28,key-29}. Although the quantum OTOCs with
imaginary Lyapunov exponent reveals the spreadability of the baryonic
wave function instead of the chaos of the system, it displays that
the baryonic status with maximum modulus of imaginary Lyapunov coefficient
(i.e. vanished theta angle) is stable which agrees with \cite{key-21,key-28,key-29}.
Moreover, we also attempt to extend our analysis with sufficiently
large density of instanton. In this case, the baryonic phase becomes
unstable which leads to a real Lyapunov exponent in OTOC at large
time. Altogether, we conclude that in the D0-D4/D8 model, real Lyapunov
exponent indicates the instability of baryonic phase while imaginary
Lyapunov exponent in OTOC indicates its metastability. And the stable
position of the baryonic phase is indicated by the maximum modulus
of the imaginary Lyapunov coefficient. Therefore, it seems the Lyapunov
coefficient as an order parameter in OTOC may detect some properties
of the QCD phase structure somehow. Besides, we evaluate the classical
limit of the OTOC whose behavior is quite different from its quantum
version. Remarkably, the instantonic dependence or theta-dependence
is less clear in the $y$ part of the classical OTOC which is quite
different from the quantum OTOCs, thus it seemingly agrees with that
the instantonic dependence or theta-dependence basically comes from
the quantum features of QCD as it is usually discussed in the framework
of QFT with instantons \cite{key-R1,key-R2,key-R3}. Overall, this
work suggests that we can use OTOC as a tool to detect the some features
of QCD phase structure.

\section*{Acknowledgements}

This work is supported by the National Natural Science Foundation
of China (NSFC) under Grant No. 12005033, the Fundamental Research
Funds for the Central Universities under Grant No. 3132024192.

\section*{Appendix: The calculation of matrix element in the OTOC}

In this appendix, we will outline the calculation of matrix element
$x_{mn}$ (\ref{eq:23}) presented in the OTOC. As the Hamiltonian
$H_{y}$ in (\ref{eq:3}) for the Skyrmion is defined in $n+1$ dimensional
moduli space, let us start with the $n+1$ dimensional Euclidean space
$\mathbb{R}^{n+1}$ parametrized by $y_{I}=\left\{ y_{1},y_{2},...y_{n+1}\right\} $
as the Cartesian coordinates. Since $H_{y}$ only depends on the radial
coordinates $\rho$ in the moduli space, we consider the spherical
coordinates $\left\{ \rho,\theta_{1},...,\theta_{n-1},\theta_{n}\right\} $
with the following coordinate transformation

\begin{equation}
y_{I}=\rho\cos\theta_{I-1}\prod_{J=I}^{n}\sin\theta_{j},\ \rho^{2}=\sum_{I=1}^{n+1}y_{I}^{2},\tag{A-1}
\end{equation}
where $\theta_{0}=0$. Thus the Euclidean metric on $\mathbb{R}^{n+1}$
is given as,

\begin{equation}
ds^{2}=\delta_{IJ}dy_{I}dy_{J}=d\rho^{2}+\rho^{2}d\Omega_{n}^{2},\tag{A-2}
\end{equation}
where $d\Omega_{n}^{2}$ represents the angular differential on a
unit $S^{n}$ satisfying the recurrence relation,

\begin{equation}
d\Omega_{n}^{2}=d\theta_{n}^{2}+\sin^{2}\theta_{n}d\Omega_{n-1}^{2}.\tag{A-3}
\end{equation}
Therefore, $d\Omega_{n}^{2}$ can be rewritten as,

\begin{equation}
d\Omega_{n}^{2}=\sum_{I=1}^{n}\left(d\theta_{I}^{2}\prod_{J=I+1}^{n}\sin^{2}\theta_{J}\right),\tag{A-4}
\end{equation}
and the volume element $dV_{n+1}$ of $\mathbb{R}^{n+1}$ reads as

\begin{align}
dV_{n+1} & =\sqrt{g}d\rho d\theta_{n}d\theta_{n-1}...d\theta_{1}=\rho^{n}d\rho\prod_{J=1}^{n}\sin^{J-1}\theta_{J}d\theta_{J}\equiv\rho^{n}d\rho dV_{S^{n}}.\tag{A-5}
\end{align}
Recall the spherical harmonic function $\mathcal{Y}_{l_{n},l_{n-1},...l_{1}}\left(\theta_{n},\theta_{n-1},...\theta_{1}\right)$
on $S^{n}$ given in (\ref{eq:10}), one can verify $\mathcal{Y}_{l_{n},l_{n-1},...l_{1}}\left(\theta_{n},\theta_{n-1},...\theta_{1}\right)$
is the eigenfunction of Laplacian operator $\nabla_{S^{n}}^{2}$ satisfying

\begin{equation}
\nabla_{S^{n}}^{2}\mathcal{Y}_{l_{n},l_{n-1},...l_{1}}\left(\theta_{n},\theta_{n-1},...\theta_{1}\right)=-l_{n}\left(l_{n}+n-1\right)\mathcal{Y}_{l_{n},l_{n-1},...l_{1}}\left(\theta_{n},\theta_{n-1},...\theta_{1}\right),\tag{A-6}
\end{equation}
where the quantum number satisfies $\left|l_{1}\right|\leq l_{2}\leq l_{3}\leq...l_{n}.$
We note that the Laplacian operator $\nabla_{S^{n}}^{2}$ satisfies
the following recurrence as,

\begin{equation}
\nabla_{S^{n+1}}^{2}=\sin^{-n}\theta_{n+1}\frac{\partial}{\partial\theta_{n+1}}\sin^{n}\theta_{n+1}\frac{\partial}{\partial\theta_{n+1}}+\sin^{-2}\theta_{n+1}\nabla_{S^{n}}^{2},\tag{A-7}
\end{equation}
where $\theta_{n+1}$ is the $n+1$-th spherical coordinate on $S^{n+1}$.
Thus, use the identity for the associated Legendre polynomial $P_{l}^{m}\left(\cos\theta\right)$

\begin{equation}
\frac{l+m}{2l+1}P_{l-1}^{m}\left(\cos\theta\right)+\frac{l-m+1}{2l+1}P_{l+1}^{m}\left(\cos\theta\right)=\cos\theta P_{l}^{m}\left(\cos\theta\right),\tag{A-8}
\end{equation}
after some messy but straightforward calculations, we can obtain

\begin{align}
\cos\theta_{n}\mathcal{Y}_{l_{n},l_{n-1},...l_{1}}= & \sqrt{\frac{\left(l_{n}-l_{n-1}\right)\left(l_{n}+l_{n-1}+n-2\right)}{\left(2l_{n}+n-1\right)\left(2l_{n}+n-3\right)}}\mathcal{Y}_{l_{n}-1,l_{n-1},...l_{1}}\nonumber \\
 & +\sqrt{\frac{\left(l_{n}+l_{n-1}+n-1\right)}{\left(2l_{n}+n-1\right)}\frac{\left(l_{n}-l_{n-1}+1\right)}{\left(2l_{n}+n+1\right)}}\mathcal{Y}_{l_{n}+1,l_{n-1},...l_{1}},\tag{A-9}
\end{align}
which leads to a useful integration,

\begin{align}
Y_{l_{n}^{\prime},l_{n-1}^{\prime},...l_{1}^{\prime};l_{n},l_{n-1},...l_{1}}\equiv & \int dV_{S^{n}}\mathcal{Y}_{l_{n}^{\prime},l_{n-1}^{\prime},...l_{1}^{\prime}}^{*}\cos\theta_{n}\mathcal{Y}_{l_{n},l_{n-1},...l_{1}}\nonumber \\
= & \bigg[\sqrt{\frac{\left(l_{n}-l_{n-1}\right)\left(l_{n}+l_{n-1}+n-2\right)}{\left(2l_{n}+n-1\right)\left(2l_{n}+n-3\right)}}\delta_{l_{n}^{\prime},l_{n}-1}\nonumber \\
 & +\sqrt{\frac{\left(l_{n}+l_{n-1}+n-1\right)}{\left(2l_{n}+n-1\right)}\frac{\left(l_{n}-l_{n-1}+1\right)}{\left(2l_{n}+n+1\right)}}\delta_{l_{n}^{\prime},l_{n}+1}\bigg]\times\prod_{j=1}^{n-1}\delta_{l_{j}^{\prime},l_{j}}.\tag{A-10}\label{eq:A-10}
\end{align}

Keep these in hand, let us choose one of the coordinates $y_{I}=\left\{ y_{1},y_{2},...y_{n+1}\right\} $
to compute its matrix element $\left\langle n_{Z},n_{\rho},l_{n},l_{n-1},...l_{1}\left|y_{I}\right|n_{Z}^{\prime},n_{\rho}^{\prime},l_{n}^{\prime},l_{n-1}^{\prime},...l_{1}^{\prime}\right\rangle $.
Due to the spherical symmetry in the Hamiltonian (\ref{eq:3}), we
can choose the $n+1$-th coordinate $y_{n+1}=\rho\cos\theta_{n}$
for simplification. Note that although we use one quantum number $\left|m\right\rangle $
to denote the eigenstate of the Hamiltonian presented in (\ref{eq:3}),
we must keep in mind the eigenstate is denoted by multiple quantum
numbers $n_{Z},n_{\rho},l_{n},l_{n-1},...l_{1}$ according to (\ref{eq:7}).
Hence we can write explicitly the matrix element $x_{mk}$ presented
in (\ref{eq:23}) as,

\begin{align}
x_{mk}= & \left\langle n_{Z},n_{\rho},l_{n},l_{n-1},...l_{1}\left|y_{n+1}\right|n_{Z}^{\prime},n_{\rho}^{\prime},l_{n}^{\prime},l_{n-1}^{\prime},...l_{1}^{\prime}\right\rangle \nonumber \\
= & \left\langle n_{Z},n_{\rho},l_{n},l_{n-1},...l_{1}\left|\rho\cos\theta_{n}\right|n_{Z}^{\prime},n_{\rho}^{\prime},l_{n}^{\prime},l_{n-1}^{\prime},...l_{1}^{\prime}\right\rangle \nonumber \\
= & \delta_{n_{Z}^{\prime},n_{Z}}\rho_{n_{\rho}^{\prime},l_{n}^{\prime};n_{\rho},l_{n}}Y_{l_{n}^{\prime},l_{n-1}^{\prime},...l_{1}^{\prime};l_{n},l_{n-1},...l_{1}},\tag{A-11}
\end{align}
where we have used the normalization of the wave function given in
(\ref{eq:7}) and (\ref{eq:A-10}) with

\begin{equation}
\rho_{n_{\rho}^{\prime},l_{n}^{\prime};n_{\rho},l_{n}}=\int\rho^{n+1}d\rho\mathcal{R}_{n_{\rho}^{\prime},l_{n}^{\prime}}\mathcal{R}_{n_{\rho},l_{n}}.\tag{A-12}
\end{equation}
By defining the kernels of the useful functions as,

\begin{align}
Y_{l_{n}^{\prime},l_{n-1}^{\prime},...l_{1}^{\prime};l_{n},l_{n-1},...l_{1}}= & Y_{l_{n}^{\prime},l_{n-1}^{\prime};l_{n},l_{n-1}}^{\mathrm{ker}}\delta_{n_{Z}^{\prime},n_{Z}}\prod_{j=1}^{n-2}\delta_{l_{j}^{\prime},l_{j}},\nonumber \\
x_{n_{\rho}^{\prime},l_{n}^{\prime},l_{n-1}^{\prime};n_{\rho},l_{n},l_{n-1}}^{\mathrm{ker}}= & \rho_{n_{\rho}^{\prime},l_{n}^{\prime};n_{\rho},l_{n}}Y_{l_{n}^{\prime},l_{n-1}^{\prime};l_{n},l_{n-1}}^{\mathrm{ker}},\nonumber \\
Y_{l_{n}^{\prime},l_{n-1}^{\prime};l_{n},l_{n-1}}^{\mathrm{ker}}= & \bigg[\sqrt{\frac{\left(l_{n}-l_{n-1}\right)\left(l_{n}+l_{n-1}+n-2\right)}{\left(2l_{n}+n-1\right)\left(2l_{n}+n-3\right)}}\delta_{l_{n}^{\prime},l_{n}-1}\nonumber \\
 & +\sqrt{\frac{\left(l_{n}+l_{n-1}+n-1\right)}{\left(2l_{n}+n-1\right)}\frac{\left(l_{n}-l_{n-1}+1\right)}{\left(2l_{n}+n+1\right)}}\delta_{l_{n}^{\prime},l_{n}+1}\bigg]\delta_{l_{n-1}^{\prime},l_{n-1}},\tag{A-13}
\end{align}
the matrix element $b_{nm}$ defined in (\ref{eq:23}) can be written
explicitly as,

\begin{align}
 & b_{n_{Z},n_{\rho},l_{n},l_{n-1},...l_{1};n_{Z}^{\prime},n_{\rho}^{\prime},l_{n}^{\prime},l_{n-1}^{\prime},...l_{1}^{\prime}}\left(t\right)\nonumber \\
= & m_{\rho}\sum_{n_{\rho}^{\prime\prime},l_{n}^{\prime\prime},l_{n-1}^{\prime\prime}}\delta_{n_{Z},n_{Z}^{\prime}}x_{n_{\rho},l_{n},l_{n-1};n_{\rho}^{\prime\prime},l_{n}^{\prime\prime},l_{n-1}^{\prime\prime}}^{\mathrm{ker}}x_{n_{\rho}^{\prime\prime},l_{n}^{\prime\prime},l_{n-1}^{\prime\prime};n_{\rho}^{\prime},l_{n}^{\prime},l_{n-1}^{\prime}}^{\mathrm{ker}}\prod_{j=1}^{n-2}\delta_{l_{j},l_{j}^{\prime}}\nonumber \\
 & \times\bigg(E_{n_{Z}^{\prime},n_{\rho}^{\prime\prime},l_{n}^{\prime\prime};n_{Z}^{\prime},n_{\rho}^{\prime},l_{n}^{\prime}}e^{iE_{n_{Z},n_{\rho},l_{n};n_{Z}^{\prime},n_{\rho}^{\prime\prime},l_{n}^{\prime\prime}}t}-E_{n_{Z},n_{\rho},l_{n};n_{Z}^{\prime},n_{\rho}^{\prime\prime},l_{n}^{\prime\prime}}e^{iE_{n_{Z}^{\prime},n_{\rho}^{\prime\prime},l_{n}^{\prime\prime};n_{Z}^{\prime},n_{\rho}^{\prime},l_{n}^{\prime}}t}\bigg),\tag{A-14}
\end{align}
where $E_{n_{Z},n_{\rho},l_{n};n_{Z}^{\prime},n_{\rho}^{\prime},l_{n}^{\prime}}$
refers to the difference of the energy defined by (\ref{eq:14}) (\ref{eq:15})
with various quantum numbers $n_{Z},n_{\rho},l_{n};n_{Z}^{\prime},n_{\rho}^{\prime},l_{n}^{\prime}$.
Therefore the microcanonical OTOC $c_{n}\left(t\right)$ can be written
as,

\begin{align}
 & c_{n_{Z},n_{\rho},l_{n},l_{n-1},...l_{1}}\left(t\right)\nonumber \\
= & \sum_{n_{Z}^{\prime},n_{\rho}^{\prime},l_{n}^{\prime},l_{n-1}^{\prime},...l_{1}^{\prime}}b_{n_{Z},n_{\rho},l_{n},l_{n-1},...l_{1};n_{Z}^{\prime},n_{\rho}^{\prime},l_{n}^{\prime},l_{n-1}^{\prime},...l_{1}^{\prime}}^{*}b_{n_{Z},n_{\rho},l_{n},l_{n-1},...l_{1};n_{Z}^{\prime},n_{\rho}^{\prime},l_{n}^{\prime},l_{n-1}^{\prime},...l_{1}^{\prime}}\nonumber \\
= & m_{\rho}^{2}\sum_{n_{\rho}^{\prime},l_{n}^{\prime},l_{n-1}^{\prime};n_{\rho}^{\prime\prime},l_{n}^{\prime\prime},l_{n-1}^{\prime\prime};n_{\rho}^{\prime\prime\prime},l_{n}^{\prime\prime\prime},l_{n-1}^{\prime\prime\prime}}\nonumber \\
 & \bigg[x_{n_{\rho},l_{n},l_{n-1};n_{\rho}^{\prime\prime},l_{n}^{\prime\prime},l_{n-1}^{\prime\prime}}^{\mathrm{ker}}x_{n_{\rho}^{\prime\prime},l_{n}^{\prime\prime},l_{n-1}^{\prime\prime};n_{\rho}^{\prime},l_{n}^{\prime},l_{n-1}^{\prime}}^{\mathrm{ker}}x_{n_{\rho},l_{n},l_{n-1};n_{\rho}^{\prime\prime\prime},l_{n}^{\prime\prime\prime},l_{n-1}^{\prime\prime\prime}}^{\mathrm{ker}}x_{n_{\rho}^{\prime\prime\prime},l_{n}^{\prime\prime\prime},l_{n-1}^{\prime\prime\prime};n_{\rho}^{\prime},l_{n}^{\prime},l_{n-1}^{\prime}}^{\mathrm{ker}}\nonumber \\
 & \times\bigg(E_{n_{\rho}^{\prime\prime},l_{n}^{\prime\prime};n_{\rho}^{\prime},l_{n}^{\prime}}e^{-iE_{n_{\rho},l_{n};n_{\rho}^{\prime\prime},l_{n}^{\prime\prime}}t}-E_{n_{\rho},l_{n};n_{\rho}^{\prime\prime},l_{n}^{\prime\prime}}e^{-iE_{n_{\rho}^{\prime\prime},l_{n}^{\prime\prime};n_{\rho}^{\prime},l_{n}^{\prime}}t}\bigg)\nonumber \\
 & \times\bigg(E_{n_{\rho}^{\prime\prime\prime},l_{n}^{\prime\prime\prime};n_{\rho}^{\prime},l_{n}^{\prime}}e^{iE_{n_{\rho},l_{n};n_{\rho}^{\prime\prime\prime},l_{n}^{\prime\prime\prime}}t}-E_{n_{\rho},l_{n};n_{\rho}^{\prime\prime\prime},l_{n}^{\prime\prime\prime}}e^{iE_{n_{\rho}^{\prime\prime\prime},l_{n}^{\prime\prime\prime};n_{\rho}^{\prime},l_{n}^{\prime}}t}\bigg)\bigg],\tag{A-15}
\end{align}
where
\begin{equation}
E_{n_{Z},n_{\rho},l_{n};n_{Z},n_{\rho}^{\prime},l_{n}^{\prime}}\equiv E_{n_{\rho},l_{n}}-E_{n_{\rho}^{\prime},l_{n}^{\prime}}=E_{n_{\rho},l_{n};n_{\rho}^{\prime},l_{n}^{\prime}}.\tag{A-16}
\end{equation}
Altogether, the OTOC is able to be evaluated numerically once the
kernel of the matrix elements $x_{n_{\rho}^{\prime\prime},l_{n}^{\prime\prime},l_{n-1}^{\prime\prime};n_{\rho}^{\prime},l_{n}^{\prime},l_{n-1}^{\prime}}^{\mathrm{ker}}$
is obtained. In addition, the thermal OTOC can be obtained by using
(\ref{eq:16}) as,

\begin{equation}
C_{T}\left(t\right)=\frac{1}{\mathcal{Z}}\sum_{n_{Z},n_{\rho},l_{n},l_{n-1}...l_{1}}c_{n_{Z},n_{\rho},l_{n},l_{n-1},...l_{1}}\left(t\right)e^{-\frac{E_{n_{Z},n_{\rho},l_{n}}}{T}},\tag{A-17}
\end{equation}
where the partition function is given by

\begin{equation}
\mathcal{Z}=\sum_{n_{Z},n_{\rho},l_{n},l_{n-1}...l_{1}}e^{-\frac{E_{n_{Z},n_{\rho},l_{n}}}{T}}=\sum_{n_{Z},n_{\rho},l_{n}}n_{\mathrm{d}}e^{-\frac{E_{n_{Z},n_{\rho},l_{n}}}{T}},\tag{A-18}
\end{equation}
and $n_{\mathrm{d}}$ is the degeneracy number in (\ref{eq:25}) for
a given energy $E_{n_{Z},n_{\rho},l_{n}}$.


\begin{thebibliography}{10}
\bibitem{key-1} J. Maldacena, S Shenke, D. Stanford, ``A bound on
chaos'', JHEP 08 (2016) 106, arXiv:1503.01409.

\bibitem{key-2} A. Larkin, Y. Ovchinnikov, ``Quasiclassical method
in the theory of superconductivity'', J. Exp. Theor. Phys. 28, 1200--1205
(1969).

\bibitem{key-3} K. Hashimoto, K. Murata, R.Yoshii, ``Out-of-time-order
correlators in quantum mechanics'', JHEP 10 (2017) 138, arXiv:1703.09435.

\bibitem{key-4} K. Hashimoto, K. Murata, K.Yoshida, ``Chaos in chiral
condensates in gauge theories'', Phys.Rev.Lett. 117 (2016) 23, 231602,
arXiv:1605.08124.

\bibitem{key-5} T. Akutagawa, K. Hashimoto, T. Sasaki, R. Watanabe,
``Out-of-time-order correlator in coupled harmonic oscillators'',
JHEP 08 (2020) 013, arXiv:2004.04381.

\bibitem{key-6}J. Maldacena, ``The Large N limit of superconformal
field theories and supergravity'', Adv.Theor.Math.Phys. 2 (1998)
231-252, arXiv:hep-th/9711200.

\bibitem{key-7} O. Aharony, S. Gubser, J. Maldacena, H. Ooguri, Y.
Oz, ``Large N field theories, string theory and gravity'', Phys.
Rept. 323 (2000) 183, arXiv:hep-th/9905111.

\bibitem{key-8}S. Shenker, D. Stanford, ``Black holes and the butterfly
effect'', JHEP 03 (2014) 067, arXiv:1306.0622.

\bibitem{key-9} S. Shenker, D. Stanford, ``Multiple Shocks'', JHEP
12 (2014) 046, arXiv:1312.3296.

\bibitem{key-10} S. Leichenauer, ``Disrupting Entanglement of Black
Holes'', Phys.Rev.D 90 (2014) 4, 046009, arXiv:1405.7365.

\bibitem{key-11} S. H. Shenker, D. Stanford, ``Stringy effects in
scrambling'', JHEP 05 (2015) 132, arXiv:1412.6087.

\bibitem{key-12} S. Jackson, L. McGough, H. Verlinde, ``Conformal
Bootstrap, Universality and Gravitational Scattering'', Nucl.Phys.B
901 (2015) 382-42, arXiv:1412.5205.

\bibitem{key-13} J. Polchinski, ``Chaos in the black hole S-matrix'',
arXiv:1505.08108.

\bibitem{key-14}S. Sachdev, J. Ye, ``Gapless spin fluid ground state
in a random, quantum Heisenberg magnet'', Phys.Rev.Lett. 70 (1993)
3339, arXiv:cond-mat/9212030.

\bibitem{key-15} A. Kitaev, talks given at KITP, April and May 2015.

\bibitem{key-16} D. J. Gross, V. Rosenhaus, ``A Generalization of
Sachdev-Ye-Kitaev'', JHEP 02 (2017) 093, arXiv:1610.01569.

\bibitem{key-17} E. Witten, ``An SYK-Like Model Without Disorder'',
J.Phys.A 52 (2019) 47, 474002, arXIv:1610.09758.

\bibitem{key-18} T. Nishinaka, S. Terashima, ``A note on Sachdev--Ye--Kitaev
like model without random coupling'', Nucl.Phys.B 926 (2018) 321-334,
arXiv:1611.10290.

\bibitem{key-19} K. Suzuki, D0 - D4 system and QCD(3+1), Phys.Rev.D
63 (2001) 084011, arXiv:hep-th/0001057.

\bibitem{key-20} S. Seki, S. Sin, \textquotedblleft A New Model of
Holographic QCD and Chiral Condensate in Dense Matter\textquotedblright ,
JHEP 10 (2013) 223 , arXiv:1304.7097.

\bibitem{key-21} L. Bartolini, F. Bigazzi, S. Bolognesi, A. Cotrone,
A. Manenti, ``Theta dependence in Holographic QCD'', JHEP 02 (2017)
029, arXiv:1611.00048.

\bibitem{key-22} C. Wu, Z. Xiao, D. Zhou, \textquotedblleft Sakai-Sugimoto
model in D0-D4 background\textquotedblright , Phys.Rev.D 88 (2013)
2, 026016 , arXiv:1304.2111.

\bibitem{key-23} T. Sakai, S. Sugimoto, \textquotedblleft Low energy
hadron physics in holographic QCD\textquotedblright , Prog.Theor.Phys.
113 (2005) 843-882 , arXiv:hep-th/0412141.

\bibitem{key-24} E. Witten, \textquotedblleft Anti-de Sitter Space,
Thermal Phase Transition, And Confinement In Gauge Theories\textquotedblright ,
Adv.Theor.Math.Phys. 2 (1998) 505-532, arXiv:hep-th/9803131.

\bibitem{key-25} T. Sakai, S. Sugimoto, \textquotedblleft More on
a holographic dual of QCD\textquotedblright , Prog.Theor.Phys. 114
(2005) 1083-1118, arXiv:hep-th/0507073.

\bibitem{key-26} E. Kiritsis, ``String theory in a nutshell'',
Princeton University Press (2019).

\bibitem{key-27} S. Li, X. Zhang, ``The D4/D8 Model and Holographic
QCD'', Symmetry 15 (2023) 6, 1213, arXiv:2304.10826.

\bibitem{key-28} W. Cai, C. Wu, Z. Xiao, ``Baryons in the Sakai-Sugimoto
model in the D0-D4 background'', Phys.Rev.D 90 (2014) 10, 106001,
arXiv:1410.5549.

\bibitem{key-29} S. Li, T. Jia, ``Matrix model and Holographic Baryons
in the D0-D4 background'', Phys.Rev.D 92 (2015) 4, 046007, arXiv:1506.00068.

\bibitem{key-R1} T.Schäfer, E. Shuryak, ``Instantons in QCD'',
Rev.Mod.Phys. 70 (1998) 323-426, arXiv:hep-ph/9610451.

\bibitem{key-R2} D. Gross, R. Pisarski, L. Yaffe, ``QCD and Instantons
at Finite Temperature'', Rev.Mod.Phys. 53 (1981) 43.

\bibitem{key-R3} D. Diakonov, N. Gromov, V. Petrov, S. Slizovskiy,
``Quantum weights of dyons and of instantons with nontrivial holonomy'',
Phys.Rev.D 70 (2004) 036003, arXiv:hep-th/0404042.

\bibitem{key-30} M. D\textquoteright Elia, F. Negro, \textquotedblleft Theta
dependence of the deconfinement temperature in Yang-Mills theories\textquotedblright ,
Phys.Rev.Lett. 109 (2012) 072001, arXiv:1205.0538.

\bibitem{key-31} M. D\textquoteright Elia, F. Negro, \textquotedblleft Phase
diagram of Yang-Mills theories in the presence of a theta term\textquotedblright ,
Phys.Rev.D 88 (2013) 3, 034503, arXiv:1306.2919.

\bibitem{key-32} L. Debbio, G. Manca, H. Panagopoulos, A. Skouroupathis,
E. Vicari, \textquotedblleft Theta-dependence of the spectrum of SU(N)
gauge theories\textquotedblright , JHEP 06 (2006) 005, arXiv:hep-th/0603041.

\bibitem{key-33} E. Witten, \textquotedblleft Theta Dependence In
The Large N Limit Of Four-Dimensional Gauge Theories\textquotedblright ,
Phys.Rev.Lett. 81 (1998) 2862-2865, arXiv:hep-th/9807109.

\bibitem{key-34} K. Buckley, T. Fugleberg, A. Zhitnitsky, \textquotedblleft Can
Induced Theta Vacua be Created in Heavy Ion Collisions?\textquotedblright ,
Phys.Rev.Lett. 84 (2000) 4814-4817, arXiv:hep-ph/9910229.

\bibitem{key-35} STAR Collaboration, ``Search for the chiral magnetic
effect with isobar collisions at $\sqrt{s_{NN}}=200$ '' by the STAR
Collaboration at the BNL Relativistic Heavy Ion Collider, Phys.Rev.C
105 (2022) 1, 014901, arXiv:2109.00131.

\bibitem{key-36} E. Vicari, H. Panagopoulos, \textquotedblleft Theta
dependence of SU(N) gauge theories in the presence of a topological
term\textquotedblright , Phys.Rept. 470 (2009) 93-150, arXiv:0803.1593.

\bibitem{key-37} E. Witten, \textquotedblleft Baryons And Branes
In Anti de Sitter Space\textquotedblright , JHEP 07 (1998) 006, arXiv:hep-th/9805112. 

\bibitem{key-38} D. Gross, H. Ooguri, \textquotedblleft Aspects of
Large N Gauge Theory Dynamics as Seen by String Theory\textquotedblright ,
Phys.Rev.D 58 (1998) 106002, arXiv:hep-th/9805129.

\bibitem{key-39} D. Tong, ``TASI lectures on solitons: Instantons,
monopoles, vortices and kinks'', TASI 2005, arXiv:hep-th/0509216.

\bibitem{key-40} H. Hata, T. Sakai, S. Sugimoto, S. Yamato, ``Baryons
from instantons in holographic QCD'', Prog.Theor.Phys. 117 (2007)
1157, arXiv:hep-th/0701280.

\bibitem{key-41} H. Hata, M.Murata, ``Baryons and the Chern-Simons
term in holographic QCD with three flavors'', Prog.Theor.Phys. 119
(2008) 461-490, arXiv:0710.2579.

\bibitem{key-42} Si-wen Li, Hao-qian Li, Yi-peng Zhang, ``The worldvolume
fermion as baryon in holographic QCD with a theta angle'', arXiv:2402.01197.

\bibitem{key-43} Si-wen Li, Yi-peng Zhang, Hao-qian Li, ``Out-of-time-order
correlators of Skyrmion as baryon in holographic QCD'', arXiv:2401.04421.

\end{thebibliography}
\end{document}